\documentclass[
twocolumn,prl,
showpacs,
nofootinbib,
amsmath,amssymb,
superscriptaddress]{revtex4-1}

\usepackage{float}
\usepackage{graphicx}
\usepackage{dcolumn}
\usepackage{bm}
\usepackage{hyperref}
\usepackage{ulem} 
\usepackage{subfig}

\newcommand{\be}{\begin{equation}}
\newcommand{\ee}{\end{equation}}
\newcommand{\bi}{\begin{itemize}}
\newcommand{\ei}{\end{itemize}}
\newcommand{\bea}{\begin{eqnarray}}
\newcommand{\eea}{\end{eqnarray}}

\def\gw#1{gravitational wave#1 (GW#1)\gdef\gw{GW}}
\def\snr#1{signal-to-noise ratio#1 (SNR#1)\gdef\snr{SNR}}
\def\bh#1{black hole#1 (BH#1)\gdef\bh{BH}}
\def\bbh#1{binary black hole#1  (BBH#1)\gdef\bbh{BBH}}
\def\nr#1{numerical relativity#1 (NR#1)\gdef\nr{NR}}
\def\gr#1{general relativity#1 (GR#1)\gdef\gr{GR}}
\def\qnm#1{quasi-normal mode#1  (QNM#1)\gdef\qnm{QNM}}
\def\fpeak#1{the instantaneous frequency at maximum amplitude#1  ($\omega_{peak}$#1)\gdef\fpeak{$\omega_{peak}$}}
\def\fdotpeak#1{the derivative of the instantaneous frequency at maximum amplitude#1  ($\dot{\omega}_{peak}$#1)\gdef\fdotpeak{$\dot{\omega}_{peak}$}}

\def\fpeaknr#1{the dimensionless instantaneous frequency at maximum amplitude#1  ($\hat{\omega}_{peak}$#1)\gdef\fpeaknr{$\hat{\omega}_{peak}$}}

\def\fdotpeaknr#1{the derivative of the dimensionless instantaneous frequency at maximum amplitude#1  ($\hat{\dot{\omega}}_{peak}$#1)\gdef\fdotpeaknr{$\hat{\dot{\omega}}_{peak}$}}

\def\cm#1{the chirp mass#1 ($\mathcal{M}$#1)\gdef\cm{$\mathcal{M}$}}
\def\af#1{the dimensionless remnant spin#1 ($a_{f}$#1)\gdef\af{$a_{f}$}}

\def\qnmfreq#1{frequency#1  ($\omega_{qnm}$#1)\gdef\qnmfreq{$\omega_{qnm}$}}
\def\qnmdecay#1{decay time#1  ($\tau_{qnm}$#1)\gdef\qnmdecay{$\tau_{qnm}$}}

\def\cwb#1{Coherent WaveBurst#1 (cWB#1)\gdef\cwb{cWB}}

\usepackage{color}

\usepackage{soul}

\begin{document}

\title{Measuring Spin of the Remnant Black Hole from Maximum Amplitude}

\affiliation{Center for Relativistic Astrophysics and School of Physics, Georgia Institute of Technology, Atlanta, GA 30332}
\affiliation{Monash Centre for Astrophysics, School of Physics and Astronomy, Monash University, VIC 3800, Australia}
\affiliation {OzGrav: The ARC Centre of Excellence for Gravitational-Wave Discovery, Clayton, VIC 3800, Australia }

\author{Deborah Ferguson$^1$}\noaffiliation
\author{Sudarshan Ghonge$^1$}\noaffiliation
\author{James A. Clark$^1$}\noaffiliation
\author{Juan Calderon Bustillo$^{2,3}$}\noaffiliation
\author{Pablo Laguna$^1$}\noaffiliation
\author{Deirdre Shoemaker $^1$}\noaffiliation

\preprint{LIGO-xxxxx}

\begin{abstract}

Gravitational waves emitted during the merger of two black holes carry information about the remnant black hole, namely its mass and spin.  This information is typically found from the ringdown radiation as the black hole settles to a final state.
We find that the remnant black hole spin is already known at the peak amplitude of the gravitational wave strain.  Using this knowledge, we present a new  method for measuring the final spin that is template independent, using only the chirp mass, the instantaneous frequency of the strain and its derivative at maximum amplitude, all template  independent.

\end{abstract}

\maketitle

\section{Introduction} \label{intro}
The advent of \gw{} astronomy has granted us the opportunity to observationally study compact binary coalescences. During the course
of the first two observing runs, LIGO \cite{TheLIGOScientific:2014jea} and Virgo \cite{TheVirgo:2014hva}  detected \gw{s} from a total of ten coalescing \bbh{s} and one 
binary neutron star~\cite{2018arXiv181112907T,PhysRevLett.119.161101}.   These systems have hinted at the population properties of \bbh{s} such as the distributions of mass, spin and redshifts~\cite{2018arXiv181112940T}, and have placed \gw{} observations into the new era of multi-messenger astronomy~\cite{PhysRevLett.119.161101,Meszaros:2019xej}.

In the few years since the first detection of \gw{s}~\cite{PhysRevLett.116.061102}, we have learned a tremendous amount about the parameter space of stellar-mass \bh{s}~\cite{2018arXiv181112940T}. 
Each stage of the coalescence provides information about the \bbh{} system; this study focuses on the parameters describing the remnant \bh{}. The product of a \bbh{} merger is a perturbed \bh{} that emits ringdown radiation as it settles to a Kerr \bh{}. This process provides fundamental information to understand gravity in its most extreme regime. Perturbation theory tells a compelling story about how perturbed \bh{s}, like the remnant of a \bbh{} merger, lose the information about the disturbance, often called hair, in the form of \gw{s} \cite{PhysRevD.1.2870}.
Perturbed \bh{s} ring down or emit \gw{s} with a \qnmfreq{} and \qnmdecay{} characterized by the \bh{} mass and spin~\cite{1971ApJ...170L.105P}, 
providing the means to determine the remnant \bh{} parameters upon the detection of \gw{s}. 

The \gw{} during this ringdown phase is generally represented as the sum of \qnm{s}, each expressible as a damped sinusoid with its own 
\qnmfreq{} and \qnmdecay{}, fixed by the mass and spin of the final \bh{} \cite{1999JMP....40..980N,1999LRR.....2....2K, 2009CQGra..26p3001B}. 
The Echeverria formulas~\cite{PhysRevD.40.3194}
provide relationships to determine the \bh{} mass and spin from \qnmfreq{} and \qnmdecay{} using spheroidal harmonics.

There have been attempts to measure \qnmfreq{} and \qnmdecay{} of the ringdown~\cite{Abbott:2009km,Goggin_2006,Berti:2007zu,Carullo:2019flw,PhysRevLett.118.161101,Nakano:2018vay,PhysRevD.97.124069}  and as the detectors improve in sensitivity, this will become more viable. One commonly considered method is to estimate the ringdown parameters by matching  directly to the 
exponentially decaying ringdown, where Ref.\cite{Carullo:2019flw} finds consistent results for GW150914 searching for damped sinusoids.    
The possibility of using \gw{s} to detect this spectrum of radiation is often referred to as \bh{} spectroscopy~\cite{Dreyer:2003bv,Berti:2005ys,PhysRevD.98.084038}.
The short duration and low-amplitude of the signal expected from stellar-mass mergers, however, 
makes this post-merger phase challenging to detect, which is further compounded by the reliance upon knowing when ringdown begins~\cite{Bhagwat:2017tkm,PhysRevD.98.104020}. 

%

Due to these challenges, current approaches~\cite{TheLIGOScientific:2016src,Healy:2014yta,Rezzolla:2007rd} to estimate the spin of the  final \bh{} match the data to theoretical models of the inspiral.  Fortunately, \nr{}  provides the map from initial to final parameters~\cite{2017PhRvD..95b4037H, 2017PhRvD..95f4024J, 2016ApJ...825L..19H} that are used to estimate the final spin.    For systems with many cycles of inspiral, this method can predict the remnant spin with precision, assuming \gr{}.  It is desirable to obtain the remnant spin independently of matched filtering of either the inspiral or ringdown in order to perform  tests of \gr{}~\cite{TheLIGOScientific:2016src,LIGOScientific:2019fpa,Ghosh:2017gfp,Ghosh:2016qgn}.  One can also perform tests of \gr{} directly from the  peak frequency~\cite{Carullo:2018gah}.  

With the goal of avoiding the use of the exponentially decaying ringdown,
we propose a method of determining the final spin that takes advantage of the
higher amplitude at the merger of two \bh{s}.
The method proposed here builds on earlier work by Healy \textit{et al}~\cite{2014CQGra..31u2001H} which
connected the instantaneous frequency of the \gw{} at peak amplitude to \qnmfreq{} and \qnmdecay{} of the 
ringdown.  While it is not obvious that such a relationship should exist, there have been hints of the merged black hole entering a perturbative regime as early as the peak amplitude~\cite{Buonanno:2006ui,PhysRevLett.109.141102,2014CQGra..31u2001H,Giesler:2019uxc} with the radiation near the peak amplitude of the strain being described by \qnm{s} that include the overtones.  In this paper, we find that the spin of the remnant black hole is already known at the peak amplitude.

Inspired by the results of Healy \textit{et al}, we create a map linking \fpeak{}, \fdotpeak{}, and \cm{} to \af{}. 
One advantage of this method is that all measurements involved, \fpeak{}, \fdotpeak{}, and \cm{}, are independent of fitting the data to a model waveform.   Furthermore, \cm{} has the advantage of needing only a few pre-peak cycles  to obtain a good measurement using a well known gravitational-wave algorithm, \cwb{}~\cite{Tiwari:2015bda}.
In the following we: a) demonstrate a tight relation between the frequency properties measured at peak and the spin of the final \bh{} and b) develop an algorithm to exploit this relationship on \gw{} observations.  

In the Methodology section, we describe the \nr{} data used to derive a 
connection from \fpeak{}, \fdotpeak{}, \cm{} 
to \af{} and discuss the 
associated errors. In the Final Spin  section, we examine the viability of the relationship as a form of parameter estimation with noisy data.  Finally, we summarize our findings in the Conclusions section.

\section{Methodology}
\label{NRfits}

\subsection{NR Catalog and Errors} \label{catalog}
The relationships found in this paper are based upon the use of 112 \nr{} simulations provided by the Georgia Tech waveform catalog, 47 of which
are nonspinning and 65 of which are aligned spin, with mass ratios $1 < q < 10$ 
\cite{2016CQGra..33t4001J}. The Georgia Tech waveforms are produced using the {\tt MAYA} code \cite{2007CQGra..24S..33H, 2007PhRvD..76h4020V, PhysRevLett.103.131101, PhysRevD.88.024040},
a branch of the {\tt Einstein Toolkit }~\cite{Loffler:2011ay}, a \nr{} code built upon {\tt Cactus} with mesh refinement from {\tt Carpet}~\cite{Schnetter:2003rb} with the addition of thorns to calculate various quantities during the simulation including an apparent horizon solver~\cite{Thornburg:2003sf}.

We create a map from \fpeak{}, \fdotpeak{}, \cm{} to \af{}.
As will be described in subsection ``Fitting to final spin,'' this equates to a mapping from
\fpeaknr{}, \fdotpeaknr{}, and the symmetric mass ratio ($\eta$) to \af{}.


In order to create this mapping, \fpeaknr{}, \fdotpeaknr{}, and \af{} are obtained from the \nr{} simulation data. 
In this paper we use the strain, $h(t)$, for ease of working with the \gw{} detectors, given   
\[
h(t) = h_+(t) -i h_{\times}(t)  = \int^{t}_{-\infty}  dt' \int^{t'}_{-\infty} dt''  \psi_4(t'') \,,
\]
and computed according to~\cite{Reisswig:2010di}.
Strain is represented as a sum of  spin-weighted spherical harmonics ${}_{-2} Y_{\ell,m}$ given by 
\[
h(t,\theta,\phi) = \sum_{\ell,m} {}_{-2}Y_{\ell,m}(\theta, \phi) h_{ \ell,m}(t)\,,
\]
where $h_{\ell, m}$ are excited depending on the inspiral parameters and the binary's orientation with respect to the observer.
In aligned spin scenarios and face on orientations, the $\ell=2$, m=2 mode dominates the signal; and, therefore, this study uses only the $\ell=2$,  $m = 2$ mode \cite{Bustillo:2016gid, Bustillo:2015qty, Pekowsky:2012sr, Varma:2016dnf}.

\begin{figure}[h]
  \centering
  \includegraphics[width = 0.45\textwidth]{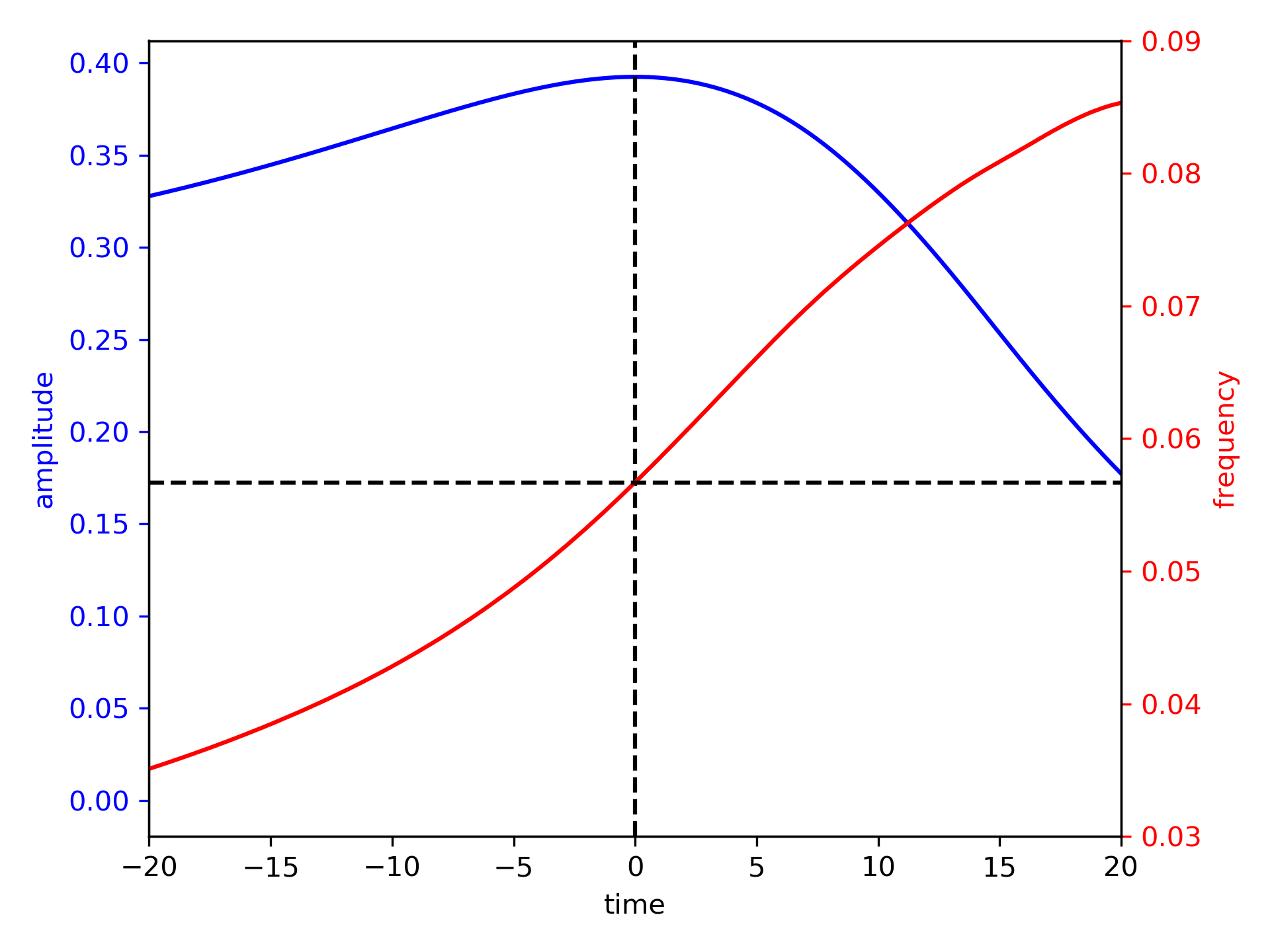}
  \caption{ The figure depicts the amplitude and the frequency during merger. The vertical dotted line denotes the time of maximum amplitude and the horizontal dotted line shows the corresponding instantaneous frequency.}
  \label{fig:peak_frequency}
\end{figure}

The \gw{} amplitude is thus $|h_{22}(t)|$, and the instantaneous frequency is found as the derivative of 
the phase, {\it i.e.} $\dot{\phi}(t)$ where $\phi(t) = arg (h_{22}(t))$. 
\fpeaknr{} and \fdotpeaknr{} are obtained simply by identifying the time at which the 
amplitude reaches a maximum and grabbing the instantaneous frequency and its time derivative at that time. This is shown visually in Fig. \ref{fig:peak_frequency}.
Note \af{} is determined from the apparent horizon of the remnant \bh{.} 

The finite spatial and temporal resolutions of \nr{} simulations
introduce systematic uncertainty 
into the estimates of frequency and spin.
By repeating each simulation at multiple 
resolutions, the error is found to be of order 0.01\% for \af{}, 1\% for \fpeaknr{}, and 1.4\% for \fdotpeaknr{}.
These uncertainties account for the spread in the fit shown in Fig. \ref{fig:fitting}.

\subsection{Fitting to final spin} \label{fitting}

With the data selected and the \nr{} errors understood, we can create a fit that connects the peak amplitude of \gw{} strain to the final BH spin. In order to create this fitting from \fpeak{}, \fdotpeak{}, and \cm{} to \af{} using \nr{} simulations, we utilize the following relationships

\begin{equation}
  \hat{\omega} \eta^{\frac{3}{5}} = \omega \mathcal{M}
\end{equation}
\begin{equation}
  \hat{\dot{\omega}} \eta^{\frac{6}{5}} = \dot{\omega} \mathcal{M}^{2} 
\end{equation}
where $\eta$ is the symmetric mass ratio defined as a function of the initial masses, $m_{1}$ and $m_{2}$:
\begin{equation}
  \eta = \frac{m_{1}m_{2}}{(m_{1}+m_{2})^{2}} \,,
\end{equation}
and \cm{} is the chirp mass expressible as:
\begin{equation}
  \mathcal{M} = \eta^{\frac{3}{5}}M = \frac{c^{3}}{G}\left( \frac{5}{96}\pi^{-\frac{8}{3}}f^{-\frac{11}{3}}\dot{f} \right) ^{\frac{3}{5}} \,.
\end{equation}
\noindent These lead us to plot the spin of the remnant \bh{} against a function of \fpeaknr{}$\eta^{\frac{3}{5}}$ and \fdotpeaknr{}$\eta^{\frac{6}{5}}$ which will take the form

\begin{equation}
  x = ln \left( \left( \hat{\omega}_{\textrm{peak}} \eta^{\frac{3}{5}} \right)^{-\frac{11}{5}} \left( \hat{\dot{\omega}}_{\textrm{peak}} \eta^{\frac{6}{5}}\right)^{\frac{4}{5}} \right).
\label{eq:x_definition}
\end{equation}

\begin{figure}
  \centering
    \includegraphics[width = 0.45\textwidth]{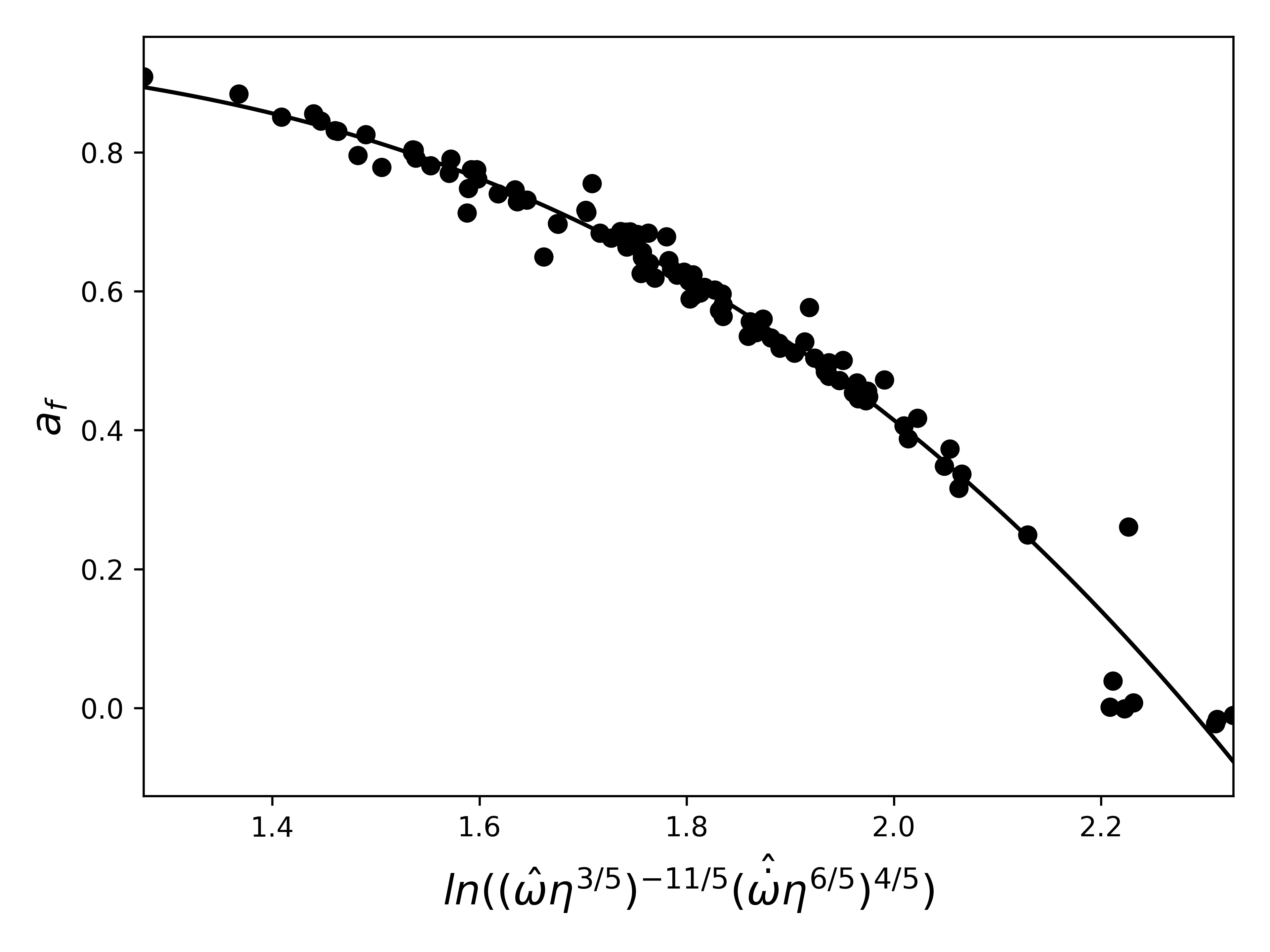}
    \caption{We plot the dimensionless spin of the remnant black hole versus a function of symmetric mass ratio, instantaneous dimensionless frequency, and its time derivative at 
    maximum strain for aligned spin numerical relativity waveforms. The solid 
    line shows the fitting relation described in the ``Fitting to final spin'' subsection. }
    \label{fig:fitting}
\end{figure}

The resulting fit is shown in Fig~\ref{fig:fitting}. 
Adopting the same functional form as Healy {\it et.al}~\cite{2014CQGra..31u2001H}, we obtain the following best fit relationship
\begin{equation}
    a_f = -0.216 x^{3} + 0.415 x^{2} -0.252 x + 0.989
  \label{eq:spin_fpeak}
\end{equation}
with an average spread of $\Delta a_f = 0.032$.

%
%


\section{Final Spin} 
\label{dataAnalysis}
Having found an \nr{} derived relationship relating \fpeak{}, \fdotpeak{}, and \cm{} to \af{,}
it's important to study how these values are obtained from real data and how precise this method will be when faced with a detection.  \cm{} is measured  by burst searches that fit the frequency evolution of the signal \cite{Tiwari:2015bda}.
By analyzing the recovered \cm{} of existing \cwb{} runs, and using the knowledge that the uncertainty scales as 1/\snr{} \cite{Vallisneri:2007ev}, we estimate that the uncertainty in \cm{} as recovered by \cwb{} is $\sim$1.5/\snr{}.
This contributes
an additional uncertainty of (126/\snr{}) \% to \af{}. For the \snr{}=100 runs we analyze in this paper, this adds an uncertainty of 1.26\% to \af{}.  

Since \gw{} detector data is noisy, we can't reliably obtain \fpeak{} and \fdotpeak{}
directly without first de-noising it. In order to reconstruct a signal out of the noise,
we use BayesWave, a search 
pipeline that relies on modeling the \gw{} as a number of sine Gaussians whose 
sum results in a coherent \gw{} signal in a detector network 
\cite{Cornish:2014kda}. 
By using this morphology-agnostic approach, the reconstructed waveform is 
robust against uncertainties which may be present in templated analyses. 
The latter model the waveform based on the time orbital evolution of Compact 
Binary Coalescences and are hence often referred to as CBC analyses~\cite{Fairhurst:2007qj}. 
BayesWave provides an independent, complementary estimate of the waveform 
morphology, and consequently avoids systematic uncertainty in the frequency 
evolution which might be present in the best fit CBC 
waveform~\cite{veitch2015parameter,2017PhRvD..95d2001M}.
In this study we analyze the waveform as reconstructed by BayesWave for the Livingston detector only.




To quantify the expected uncertainty in
the remnant spin,
we performed a systematic Monte-Carlo study whereby sets of BBH signals with 
increasing \snr{}  \cite{Parzen:2006:RES:2263383.2270618} 
were added to stationary Gaussian noise 
colored with the power spectral density of O1 era LIGO detectors.  The 
underlying waveforms for these ``injections'' were then recovered using 
BayesWave. 
For a \snr{} of 100, we injected a $h_{22}$ signal consistent with that of GW150914 in 2000 realizations of Gaussian noise and recovered \fpeak{} and \fdotpeak{} for the median waveform of each.
The value of \fpeak{} was obtained by first calculating the amplitude envelope of the median whitened waveform
(using a python implementation of the Hilbert-Huang transform \cite{doi:10.1098/rspa.1998.0193})
and then locating the time at which the amplitude is maximum. Then the median time frequency
track, outputted by BayesWave, is used to identify the frequency and the time derivative of the frequency at the given time.  
%





\begin{figure}
  \centering
  \includegraphics[width = 0.45\textwidth]{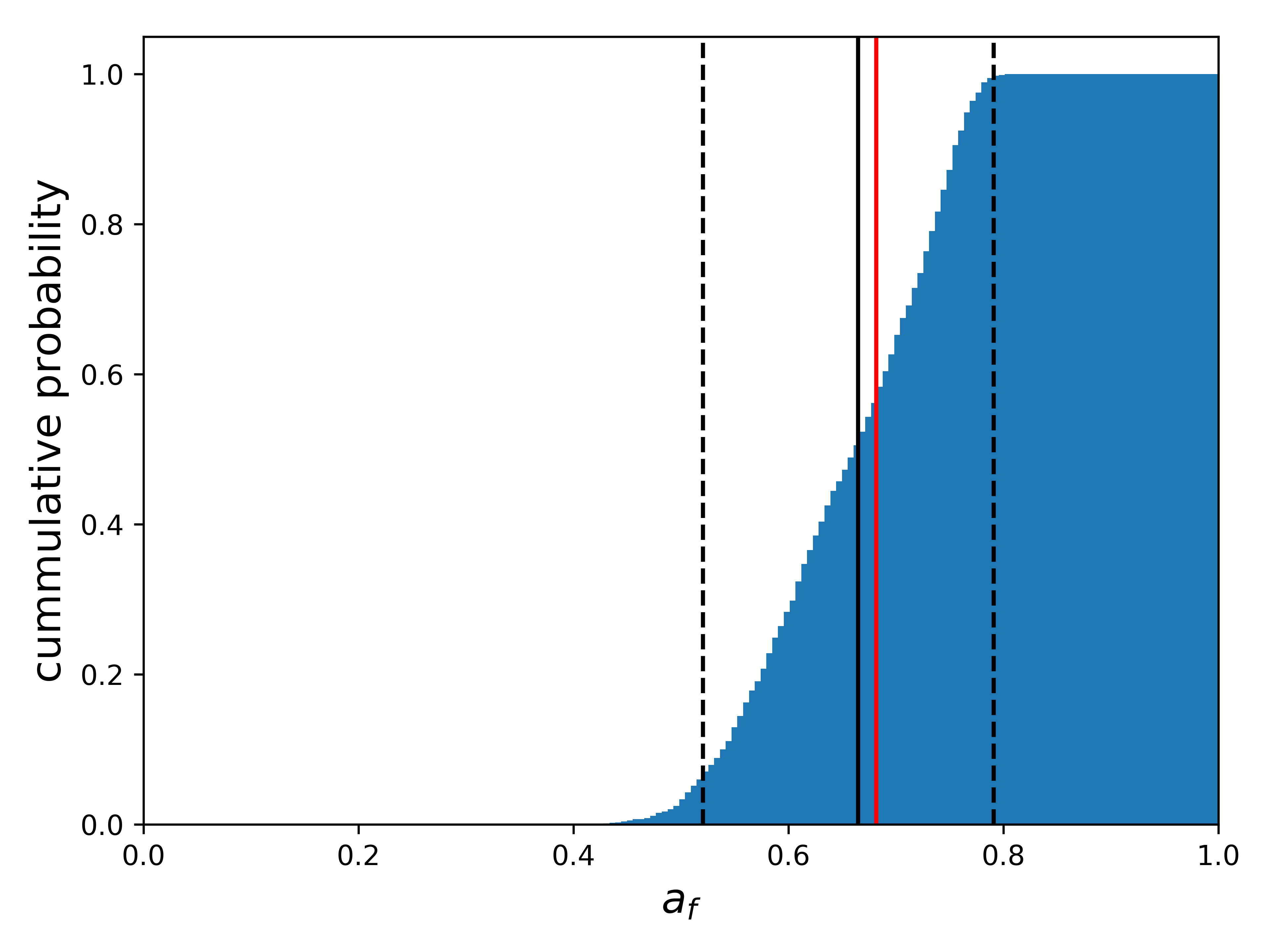}
  \caption{We plot the cumulative probability distribution of the final dimensionless spin obtained for a GW150914-like signal
    injected into noise and recovered using BayesWave with \snr{} 100. The solid black line shows the median
    recovered spin and the dotted black lines show the 90\% confidence interval. The solid red line shows the
    true spin. }
  \label{fig:histogram}
\end{figure}

Fig \ref{fig:histogram} shows the cumulative probability distribution of the estimated \af{} for our 2000 injections.
The solid black line denotes the median, the solid red line denotes the true final spin, and
the dotted lines show the 90\% confidence interval, which is \af{}=(0.51, 0.77) for SNR of 100.

%
%

To better understand how this error scales with \snr{}, we used the same technique just described with 250 injections each for \snr{s} 40, 60, 80, and 100. The resulting medians and 90\% confidence intervals are shown in Table \ref{tbl:snr}.

\begin{table}[h]
  \centering
  \begin{tabular}{c|c|c|c}
    SNR & median & lower 90\% confidence & upper 90\% confidence \\ 
    \hline
    40 & 0.671 & 0.437 & 0.802 \\
    60 & 0.677 & 0.484 & 0.785 \\
    80 & 0.654 & 0.497 & 0.782 \\
    100 & 0.667 & 0.510 & 0.772
  \end{tabular}
  \caption{Median and 90\% confidence values of \af{} for various \snr{s}}
  \label{tbl:snr}
\end{table}

\section{Conclusions} \label{conclusion}

This study finds that the remnant spin is known at the peak amplitude and presents a method of estimating it from the
 chirp mass, the frequency at maximum amplitude of the strain, and its derivative in an analytic relationship.   This allows us to make use of the high SNR at the peak to estimate the final spin before entering the perturbative ringdown regime. 

In order to understand the viability of this study as a parameter estimation method, we analyzed the
 distribution of the remnant spin obtained via recovering the waveform of a GW150914-like signal with increasing \snr{s} from 40 to 100.
 We demonstrate that we can reliably place bounds on the spin of the remnant \bh{} using information found near the peak amplitude  when the signal is dominated by the $\ell=2$, $m=2$ mode.

 
Our method avoids the usage of \bbh{} templates,  instead obtaining \fpeak{} and \fdotpeak{} from a BayesWave reconstruction and  \cm{} from \cwb{}.   
While matched filtering methods  likely place a tighter bound on 
the remnant spin, our alternate approach is not subject to the same systematic biases due to waveform modeling present in the matched filter search.   There 
remain systematic errors due to the fit we are using to determine the final
spin from the peak amplitude.  In addition, the fitting formula is an interpolation
over a discrete set of NR templates and might change if more NR simulations are
added to the fit.


Next steps in this study will see the method  applied to all the  LIGO/VIRGO BBH 
detections with reasonable BayesWave reconstructions from O1, O2 and, soon,  O3.
It will also be interesting to see the effect of adding precessing runs to the 
fit and whether this analysis can be expanded to include higher modes.


\paragraph*{\textbf{Acknowledgements}}
PL and DS gratefully acknowledge support from the NSF grants PHY-1806580, PHY-1809572, PHY-1550461 and 1333360, XSEDE  TG-PHY120016. JCB also acknowledges support from Australian Research Council Discovery Project DP180103155. This research was also supported in part through research cyberinfrastructure resources and services provided by the Partnership for an Advanced Computing Environment (PACE) at the Georgia Institute of Technology. The authors are grateful for computational resources provided by the LIGO Laboratory and supported by National Science Foundation Grants PHY-0757058 and PHY-0823459.   We also thank Karan Jani for useful discussions. 

\bibliography{refs}

\begin{thebibliography}{60}%
\makeatletter
\providecommand \@ifxundefined [1]{%
 \@ifx{#1\undefined}
}%
\providecommand \@ifnum [1]{%
 \ifnum #1\expandafter \@firstoftwo
 \else \expandafter \@secondoftwo
 \fi
}%
\providecommand \@ifx [1]{%
 \ifx #1\expandafter \@firstoftwo
 \else \expandafter \@secondoftwo
 \fi
}%
\providecommand \natexlab [1]{#1}%
\providecommand \enquote  [1]{``#1''}%
\providecommand \bibnamefont  [1]{#1}%
\providecommand \bibfnamefont [1]{#1}%
\providecommand \citenamefont [1]{#1}%
\providecommand \href@noop [0]{\@secondoftwo}%
\providecommand \href [0]{\begingroup \@sanitize@url \@href}%
\providecommand \@href[1]{\@@startlink{#1}\@@href}%
\providecommand \@@href[1]{\endgroup#1\@@endlink}%
\providecommand \@sanitize@url [0]{\catcode `\\12\catcode `\$12\catcode
  `\&12\catcode `\#12\catcode `\^12\catcode `\_12\catcode `\%12\relax}%
\providecommand \@@startlink[1]{}%
\providecommand \@@endlink[0]{}%
\providecommand \url  [0]{\begingroup\@sanitize@url \@url }%
\providecommand \@url [1]{\endgroup\@href {#1}{\urlprefix }}%
\providecommand \urlprefix  [0]{URL }%
\providecommand \Eprint [0]{\href }%
\providecommand \doibase [0]{http://dx.doi.org/}%
\providecommand \selectlanguage [0]{\@gobble}%
\providecommand \bibinfo  [0]{\@secondoftwo}%
\providecommand \bibfield  [0]{\@secondoftwo}%
\providecommand \translation [1]{[#1]}%
\providecommand \BibitemOpen [0]{}%
\providecommand \bibitemStop [0]{}%
\providecommand \bibitemNoStop [0]{.\EOS\space}%
\providecommand \EOS [0]{\spacefactor3000\relax}%
\providecommand \BibitemShut  [1]{\csname bibitem#1\endcsname}%
\let\auto@bib@innerbib\@empty
\bibitem [{\citenamefont {Aasi}\ \emph {et~al.}(2015)\citenamefont {Aasi} \emph
  {et~al.}}]{TheLIGOScientific:2014jea}%
  \BibitemOpen
  \bibfield  {author} {\bibinfo {author} {\bibfnamefont {J.}~\bibnamefont
  {Aasi}} \emph {et~al.} (\bibinfo {collaboration} {LIGO Scientific}),\ }\href
  {\doibase 10.1088/0264-9381/32/7/074001} {\bibfield  {journal} {\bibinfo
  {journal} {Class. Quant. Grav.}\ }\textbf {\bibinfo {volume} {32}},\ \bibinfo
  {pages} {074001} (\bibinfo {year} {2015})},\ \Eprint
  {http://arxiv.org/abs/1411.4547} {arXiv:1411.4547 [gr-qc]} \BibitemShut
  {NoStop}%
\bibitem [{\citenamefont {Acernese}\ \emph {et~al.}(2015)\citenamefont
  {Acernese} \emph {et~al.}}]{TheVirgo:2014hva}%
  \BibitemOpen
  \bibfield  {author} {\bibinfo {author} {\bibfnamefont {F.}~\bibnamefont
  {Acernese}} \emph {et~al.} (\bibinfo {collaboration} {VIRGO}),\ }\href
  {\doibase 10.1088/0264-9381/32/2/024001} {\bibfield  {journal} {\bibinfo
  {journal} {Class. Quant. Grav.}\ }\textbf {\bibinfo {volume} {32}},\ \bibinfo
  {pages} {024001} (\bibinfo {year} {2015})},\ \Eprint
  {http://arxiv.org/abs/1408.3978} {arXiv:1408.3978 [gr-qc]} \BibitemShut
  {NoStop}%
\bibitem [{\citenamefont {{The LIGO Scientific Collaboration}}\ and\
  \citenamefont {{the Virgo Collaboration}}(2018)}]{2018arXiv181112907T}%
  \BibitemOpen
  \bibfield  {author} {\bibinfo {author} {\bibnamefont {{The LIGO Scientific
  Collaboration}}}\ and\ \bibinfo {author} {\bibnamefont {{the Virgo
  Collaboration}}},\ }\href@noop {} {\bibfield  {journal} {\bibinfo  {journal}
  {arXiv e-prints}\ ,\ \bibinfo {eid} {arXiv:1811.12907}} (\bibinfo {year}
  {2018})},\ \Eprint {http://arxiv.org/abs/1811.12907} {1811.12907
  [astro-ph.HE]} \BibitemShut {NoStop}%
\bibitem [{\citenamefont {Collaboration}\ and\ \citenamefont
  {Collaboration}(2017)}]{PhysRevLett.119.161101}%
  \BibitemOpen
  \bibfield  {author} {\bibinfo {author} {\bibfnamefont {L.~S.}\ \bibnamefont
  {Collaboration}}\ and\ \bibinfo {author} {\bibfnamefont {V.}~\bibnamefont
  {Collaboration}},\ }\href {\doibase 10.1103/PhysRevLett.119.161101}
  {\bibfield  {journal} {\bibinfo  {journal} {Phys. Rev. Lett.}\ }\textbf
  {\bibinfo {volume} {119}},\ \bibinfo {pages} {161101} (\bibinfo {year}
  {2017})}\BibitemShut {NoStop}%
\bibitem [{\citenamefont {{The LIGO Scientific Collaboration}}\ and\
  \citenamefont {{The Virgo Collaboration}}(2018)}]{2018arXiv181112940T}%
  \BibitemOpen
  \bibfield  {author} {\bibinfo {author} {\bibnamefont {{The LIGO Scientific
  Collaboration}}}\ and\ \bibinfo {author} {\bibnamefont {{The Virgo
  Collaboration}}},\ }\href@noop {} {\bibfield  {journal} {\bibinfo  {journal}
  {arXiv e-prints}\ ,\ \bibinfo {eid} {arXiv:1811.12940}} (\bibinfo {year}
  {2018})},\ \Eprint {http://arxiv.org/abs/1811.12940} {1811.12940
  [astro-ph.HE]} \BibitemShut {NoStop}%
\bibitem [{\citenamefont {Ma\'esz\`aros}\ \emph {et~al.}(2019)\citenamefont
  {Ma\'esz\`aros}, \citenamefont {Fox}, \citenamefont {Hanna},\ and\
  \citenamefont {Murase}}]{Meszaros:2019xej}%
  \BibitemOpen
  \bibfield  {author} {\bibinfo {author} {\bibfnamefont {P.}~\bibnamefont
  {Ma\'esz\`aros}}, \bibinfo {author} {\bibfnamefont {D.~B.}\ \bibnamefont
  {Fox}}, \bibinfo {author} {\bibfnamefont {C.}~\bibnamefont {Hanna}}, \ and\
  \bibinfo {author} {\bibfnamefont {K.}~\bibnamefont {Murase}},\ }\href@noop {}
  {\  (\bibinfo {year} {2019})},\ \Eprint {http://arxiv.org/abs/1906.10212}
  {arXiv:1906.10212 [astro-ph.HE]} \BibitemShut {NoStop}%
\bibitem [{\citenamefont {Collaboration}\ and\ \citenamefont
  {Collaboration}(2016)}]{PhysRevLett.116.061102}%
  \BibitemOpen
  \bibfield  {author} {\bibinfo {author} {\bibfnamefont {L.~S.}\ \bibnamefont
  {Collaboration}}\ and\ \bibinfo {author} {\bibfnamefont {V.}~\bibnamefont
  {Collaboration}},\ }\href {\doibase 10.1103/PhysRevLett.116.061102}
  {\bibfield  {journal} {\bibinfo  {journal} {Phys. Rev. Lett.}\ }\textbf
  {\bibinfo {volume} {116}},\ \bibinfo {pages} {061102} (\bibinfo {year}
  {2016})}\BibitemShut {NoStop}%
\bibitem [{\citenamefont {Vishveshwara}(1970)}]{PhysRevD.1.2870}%
  \BibitemOpen
  \bibfield  {author} {\bibinfo {author} {\bibfnamefont {C.~V.}\ \bibnamefont
  {Vishveshwara}},\ }\href {\doibase 10.1103/PhysRevD.1.2870} {\bibfield
  {journal} {\bibinfo  {journal} {Phys. Rev. D}\ }\textbf {\bibinfo {volume}
  {1}},\ \bibinfo {pages} {2870} (\bibinfo {year} {1970})}\BibitemShut
  {NoStop}%
\bibitem [{\citenamefont {{Press}}(1971)}]{1971ApJ...170L.105P}%
  \BibitemOpen
  \bibfield  {author} {\bibinfo {author} {\bibfnamefont {W.~H.}\ \bibnamefont
  {{Press}}},\ }\href {\doibase 10.1086/180849} {\bibfield  {journal} {\bibinfo
   {journal} {ApJ Letters}\ }\textbf {\bibinfo {volume} {170}},\ \bibinfo
  {pages} {L105} (\bibinfo {year} {1971})}\BibitemShut {NoStop}%
\bibitem [{\citenamefont {{Nollert}}\ and\ \citenamefont
  {{Price}}(1999)}]{1999JMP....40..980N}%
  \BibitemOpen
  \bibfield  {author} {\bibinfo {author} {\bibfnamefont {H.-P.}\ \bibnamefont
  {{Nollert}}}\ and\ \bibinfo {author} {\bibfnamefont {R.~H.}\ \bibnamefont
  {{Price}}},\ }\href {\doibase 10.1063/1.532698} {\bibfield  {journal}
  {\bibinfo  {journal} {Journal of Mathematical Physics}\ }\textbf {\bibinfo
  {volume} {40}},\ \bibinfo {pages} {980} (\bibinfo {year} {1999})},\ \Eprint
  {http://arxiv.org/abs/gr-qc/9810074} {arXiv:gr-qc/9810074 [gr-qc]}
  \BibitemShut {NoStop}%
\bibitem [{\citenamefont {{Kokkotas}}\ and\ \citenamefont
  {{Schmidt}}(1999)}]{1999LRR.....2....2K}%
  \BibitemOpen
  \bibfield  {author} {\bibinfo {author} {\bibfnamefont {K.~D.}\ \bibnamefont
  {{Kokkotas}}}\ and\ \bibinfo {author} {\bibfnamefont {B.~G.}\ \bibnamefont
  {{Schmidt}}},\ }\href {\doibase 10.12942/lrr-1999-2} {\bibfield  {journal}
  {\bibinfo  {journal} {Living Reviews in Relativity}\ }\textbf {\bibinfo
  {volume} {2}},\ \bibinfo {eid} {2} (\bibinfo {year} {1999})},\ \Eprint
  {http://arxiv.org/abs/gr-qc/9909058} {arXiv:gr-qc/9909058 [gr-qc]}
  \BibitemShut {NoStop}%
\bibitem [{\citenamefont {{Berti}}\ \emph {et~al.}(2009)\citenamefont
  {{Berti}}, \citenamefont {{Cardoso}},\ and\ \citenamefont
  {{Starinets}}}]{2009CQGra..26p3001B}%
  \BibitemOpen
  \bibfield  {author} {\bibinfo {author} {\bibfnamefont {E.}~\bibnamefont
  {{Berti}}}, \bibinfo {author} {\bibfnamefont {V.}~\bibnamefont {{Cardoso}}},
  \ and\ \bibinfo {author} {\bibfnamefont {A.~O.}\ \bibnamefont
  {{Starinets}}},\ }\href {\doibase 10.1088/0264-9381/26/16/163001} {\bibfield
  {journal} {\bibinfo  {journal} {Classical and Quantum Gravity}\ }\textbf
  {\bibinfo {volume} {26}},\ \bibinfo {eid} {163001} (\bibinfo {year}
  {2009})},\ \Eprint {http://arxiv.org/abs/0905.2975} {arXiv:0905.2975 [gr-qc]}
  \BibitemShut {NoStop}%
\bibitem [{\citenamefont {Echeverria}(1989)}]{PhysRevD.40.3194}%
  \BibitemOpen
  \bibfield  {author} {\bibinfo {author} {\bibfnamefont {F.}~\bibnamefont
  {Echeverria}},\ }\href {\doibase 10.1103/PhysRevD.40.3194} {\bibfield
  {journal} {\bibinfo  {journal} {Phys. Rev. D}\ }\textbf {\bibinfo {volume}
  {40}},\ \bibinfo {pages} {3194} (\bibinfo {year} {1989})}\BibitemShut
  {NoStop}%
\bibitem [{\citenamefont {Abbott}\ \emph {et~al.}(2009)\citenamefont {Abbott}
  \emph {et~al.}}]{Abbott:2009km}%
  \BibitemOpen
  \bibfield  {author} {\bibinfo {author} {\bibfnamefont {B.~P.}\ \bibnamefont
  {Abbott}} \emph {et~al.} (\bibinfo {collaboration} {LIGO Scientific}),\
  }\href {\doibase 10.1103/PhysRevD.80.062001} {\bibfield  {journal} {\bibinfo
  {journal} {Phys. Rev.}\ }\textbf {\bibinfo {volume} {D80}},\ \bibinfo {pages}
  {062001} (\bibinfo {year} {2009})},\ \Eprint {http://arxiv.org/abs/0905.1654}
  {arXiv:0905.1654 [gr-qc]} \BibitemShut {NoStop}%
\bibitem [{\citenamefont {Goggin}\ and\ \citenamefont {the LIGO
  Scientific~Collaboration}(2006)}]{Goggin_2006}%
  \BibitemOpen
  \bibfield  {author} {\bibinfo {author} {\bibfnamefont {L.~M.}\ \bibnamefont
  {Goggin}}\ and\ \bibinfo {author} {\bibnamefont {the LIGO
  Scientific~Collaboration}},\ }\href {\doibase 10.1088/0264-9381/23/19/s09}
  {\bibfield  {journal} {\bibinfo  {journal} {Classical and Quantum Gravity}\
  }\textbf {\bibinfo {volume} {23}},\ \bibinfo {pages} {S709} (\bibinfo {year}
  {2006})}\BibitemShut {NoStop}%
\bibitem [{\citenamefont {Berti}\ \emph {et~al.}(2007)\citenamefont {Berti},
  \citenamefont {Cardoso}, \citenamefont {Cardoso},\ and\ \citenamefont
  {Cavaglia}}]{Berti:2007zu}%
  \BibitemOpen
  \bibfield  {author} {\bibinfo {author} {\bibfnamefont {E.}~\bibnamefont
  {Berti}}, \bibinfo {author} {\bibfnamefont {J.}~\bibnamefont {Cardoso}},
  \bibinfo {author} {\bibfnamefont {V.}~\bibnamefont {Cardoso}}, \ and\
  \bibinfo {author} {\bibfnamefont {M.}~\bibnamefont {Cavaglia}},\ }\href
  {\doibase 10.1103/PhysRevD.76.104044} {\bibfield  {journal} {\bibinfo
  {journal} {Phys. Rev.}\ }\textbf {\bibinfo {volume} {D76}},\ \bibinfo {pages}
  {104044} (\bibinfo {year} {2007})},\ \Eprint {http://arxiv.org/abs/0707.1202}
  {arXiv:0707.1202 [gr-qc]} \BibitemShut {NoStop}%
\bibitem [{\citenamefont {Carullo}\ \emph
  {et~al.}(2019{\natexlab{a}})\citenamefont {Carullo}, \citenamefont
  {Del~Pozzo},\ and\ \citenamefont {Veitch}}]{Carullo:2019flw}%
  \BibitemOpen
  \bibfield  {author} {\bibinfo {author} {\bibfnamefont {G.}~\bibnamefont
  {Carullo}}, \bibinfo {author} {\bibfnamefont {W.}~\bibnamefont {Del~Pozzo}},
  \ and\ \bibinfo {author} {\bibfnamefont {J.}~\bibnamefont {Veitch}},\
  }\href@noop {} {\bibfield  {journal} {\bibinfo  {journal} {arXiv:1902.07527}\
  } (\bibinfo {year} {2019}{\natexlab{a}})}\BibitemShut {NoStop}%
\bibitem [{\citenamefont {Yang}\ \emph {et~al.}(2017)\citenamefont {Yang},
  \citenamefont {Yagi}, \citenamefont {Blackman}, \citenamefont {Lehner},
  \citenamefont {Paschalidis}, \citenamefont {Pretorius},\ and\ \citenamefont
  {Yunes}}]{PhysRevLett.118.161101}%
  \BibitemOpen
  \bibfield  {author} {\bibinfo {author} {\bibfnamefont {H.}~\bibnamefont
  {Yang}}, \bibinfo {author} {\bibfnamefont {K.}~\bibnamefont {Yagi}}, \bibinfo
  {author} {\bibfnamefont {J.}~\bibnamefont {Blackman}}, \bibinfo {author}
  {\bibfnamefont {L.}~\bibnamefont {Lehner}}, \bibinfo {author} {\bibfnamefont
  {V.}~\bibnamefont {Paschalidis}}, \bibinfo {author} {\bibfnamefont
  {F.}~\bibnamefont {Pretorius}}, \ and\ \bibinfo {author} {\bibfnamefont
  {N.}~\bibnamefont {Yunes}},\ }\href {\doibase 10.1103/PhysRevLett.118.161101}
  {\bibfield  {journal} {\bibinfo  {journal} {Phys. Rev. Lett.}\ }\textbf
  {\bibinfo {volume} {118}},\ \bibinfo {pages} {161101} (\bibinfo {year}
  {2017})}\BibitemShut {NoStop}%
\bibitem [{\citenamefont {Nakano}\ \emph {et~al.}(2018)\citenamefont {Nakano},
  \citenamefont {Narikawa}, \citenamefont {Oohara}, \citenamefont {Sakai},
  \citenamefont {Shinkai}, \citenamefont {Takahashi}, \citenamefont {Tanaka},
  \citenamefont {Uchikata}, \citenamefont {Yamamoto},\ and\ \citenamefont
  {Yamamoto}}]{Nakano:2018vay}%
  \BibitemOpen
  \bibfield  {author} {\bibinfo {author} {\bibfnamefont {H.}~\bibnamefont
  {Nakano}}, \bibinfo {author} {\bibfnamefont {T.}~\bibnamefont {Narikawa}},
  \bibinfo {author} {\bibfnamefont {K.-i.}\ \bibnamefont {Oohara}}, \bibinfo
  {author} {\bibfnamefont {K.}~\bibnamefont {Sakai}}, \bibinfo {author}
  {\bibfnamefont {H.-a.}\ \bibnamefont {Shinkai}}, \bibinfo {author}
  {\bibfnamefont {H.}~\bibnamefont {Takahashi}}, \bibinfo {author}
  {\bibfnamefont {T.}~\bibnamefont {Tanaka}}, \bibinfo {author} {\bibfnamefont
  {N.}~\bibnamefont {Uchikata}}, \bibinfo {author} {\bibfnamefont
  {S.}~\bibnamefont {Yamamoto}}, \ and\ \bibinfo {author} {\bibfnamefont
  {T.~S.}\ \bibnamefont {Yamamoto}},\ }\href@noop {} {\  (\bibinfo {year}
  {2018})},\ \Eprint {http://arxiv.org/abs/1811.06443} {arXiv:1811.06443
  [gr-qc]} \BibitemShut {NoStop}%
\bibitem [{\citenamefont {Cabero}\ \emph {et~al.}(2018)\citenamefont {Cabero},
  \citenamefont {Capano}, \citenamefont {Fischer-Birnholtz}, \citenamefont
  {Krishnan}, \citenamefont {Nielsen}, \citenamefont {Nitz},\ and\
  \citenamefont {Biwer}}]{PhysRevD.97.124069}%
  \BibitemOpen
  \bibfield  {author} {\bibinfo {author} {\bibfnamefont {M.}~\bibnamefont
  {Cabero}}, \bibinfo {author} {\bibfnamefont {C.~D.}\ \bibnamefont {Capano}},
  \bibinfo {author} {\bibfnamefont {O.}~\bibnamefont {Fischer-Birnholtz}},
  \bibinfo {author} {\bibfnamefont {B.}~\bibnamefont {Krishnan}}, \bibinfo
  {author} {\bibfnamefont {A.~B.}\ \bibnamefont {Nielsen}}, \bibinfo {author}
  {\bibfnamefont {A.~H.}\ \bibnamefont {Nitz}}, \ and\ \bibinfo {author}
  {\bibfnamefont {C.~M.}\ \bibnamefont {Biwer}},\ }\href {\doibase
  10.1103/PhysRevD.97.124069} {\bibfield  {journal} {\bibinfo  {journal} {Phys.
  Rev. D}\ }\textbf {\bibinfo {volume} {97}},\ \bibinfo {pages} {124069}
  (\bibinfo {year} {2018})}\BibitemShut {NoStop}%
\bibitem [{\citenamefont {Dreyer}\ \emph {et~al.}(2004)\citenamefont {Dreyer},
  \citenamefont {Kelly}, \citenamefont {Krishnan}, \citenamefont {Finn},
  \citenamefont {Garrison},\ and\ \citenamefont
  {Lopez-Aleman}}]{Dreyer:2003bv}%
  \BibitemOpen
  \bibfield  {author} {\bibinfo {author} {\bibfnamefont {O.}~\bibnamefont
  {Dreyer}}, \bibinfo {author} {\bibfnamefont {B.~J.}\ \bibnamefont {Kelly}},
  \bibinfo {author} {\bibfnamefont {B.}~\bibnamefont {Krishnan}}, \bibinfo
  {author} {\bibfnamefont {L.~S.}\ \bibnamefont {Finn}}, \bibinfo {author}
  {\bibfnamefont {D.}~\bibnamefont {Garrison}}, \ and\ \bibinfo {author}
  {\bibfnamefont {R.}~\bibnamefont {Lopez-Aleman}},\ }\href {\doibase
  10.1088/0264-9381/21/4/003} {\bibfield  {journal} {\bibinfo  {journal}
  {Class. Quant. Grav.}\ }\textbf {\bibinfo {volume} {21}},\ \bibinfo {pages}
  {787} (\bibinfo {year} {2004})},\ \Eprint
  {http://arxiv.org/abs/gr-qc/0309007} {arXiv:gr-qc/0309007 [gr-qc]}
  \BibitemShut {NoStop}%
\bibitem [{\citenamefont {Berti}\ \emph {et~al.}(2006)\citenamefont {Berti},
  \citenamefont {Cardoso},\ and\ \citenamefont {Will}}]{Berti:2005ys}%
  \BibitemOpen
  \bibfield  {author} {\bibinfo {author} {\bibfnamefont {E.}~\bibnamefont
  {Berti}}, \bibinfo {author} {\bibfnamefont {V.}~\bibnamefont {Cardoso}}, \
  and\ \bibinfo {author} {\bibfnamefont {C.~M.}\ \bibnamefont {Will}},\ }\href
  {\doibase 10.1103/PhysRevD.73.064030} {\bibfield  {journal} {\bibinfo
  {journal} {Phys. Rev.}\ }\textbf {\bibinfo {volume} {D73}},\ \bibinfo {pages}
  {064030} (\bibinfo {year} {2006})},\ \Eprint
  {http://arxiv.org/abs/gr-qc/0512160} {arXiv:gr-qc/0512160 [gr-qc]}
  \BibitemShut {NoStop}%
\bibitem [{\citenamefont {Brito}\ \emph {et~al.}(2018)\citenamefont {Brito},
  \citenamefont {Buonanno},\ and\ \citenamefont
  {Raymond}}]{PhysRevD.98.084038}%
  \BibitemOpen
  \bibfield  {author} {\bibinfo {author} {\bibfnamefont {R.}~\bibnamefont
  {Brito}}, \bibinfo {author} {\bibfnamefont {A.}~\bibnamefont {Buonanno}}, \
  and\ \bibinfo {author} {\bibfnamefont {V.}~\bibnamefont {Raymond}},\ }\href
  {\doibase 10.1103/PhysRevD.98.084038} {\bibfield  {journal} {\bibinfo
  {journal} {Phys. Rev. D}\ }\textbf {\bibinfo {volume} {98}},\ \bibinfo
  {pages} {084038} (\bibinfo {year} {2018})}\BibitemShut {NoStop}%
\bibitem [{\citenamefont {Bhagwat}\ \emph {et~al.}(2018)\citenamefont
  {Bhagwat}, \citenamefont {Okounkova}, \citenamefont {Ballmer}, \citenamefont
  {Brown}, \citenamefont {Giesler}, \citenamefont {Scheel},\ and\ \citenamefont
  {Teukolsky}}]{Bhagwat:2017tkm}%
  \BibitemOpen
  \bibfield  {author} {\bibinfo {author} {\bibfnamefont {S.}~\bibnamefont
  {Bhagwat}}, \bibinfo {author} {\bibfnamefont {M.}~\bibnamefont {Okounkova}},
  \bibinfo {author} {\bibfnamefont {S.~W.}\ \bibnamefont {Ballmer}}, \bibinfo
  {author} {\bibfnamefont {D.~A.}\ \bibnamefont {Brown}}, \bibinfo {author}
  {\bibfnamefont {M.}~\bibnamefont {Giesler}}, \bibinfo {author} {\bibfnamefont
  {M.~A.}\ \bibnamefont {Scheel}}, \ and\ \bibinfo {author} {\bibfnamefont
  {S.~A.}\ \bibnamefont {Teukolsky}},\ }\href {\doibase
  10.1103/PhysRevD.97.104065} {\bibfield  {journal} {\bibinfo  {journal} {Phys.
  Rev.}\ }\textbf {\bibinfo {volume} {D97}},\ \bibinfo {pages} {104065}
  (\bibinfo {year} {2018})},\ \Eprint {http://arxiv.org/abs/1711.00926}
  {arXiv:1711.00926 [gr-qc]} \BibitemShut {NoStop}%
\bibitem [{\citenamefont {Carullo}\ \emph {et~al.}(2018)\citenamefont
  {Carullo}, \citenamefont {van~der Schaaf}, \citenamefont {London},
  \citenamefont {Pang}, \citenamefont {Tsang}, \citenamefont {Hannuksela},
  \citenamefont {Meidam}, \citenamefont {Agathos}, \citenamefont {Samajdar},
  \citenamefont {Ghosh}, \citenamefont {Li}, \citenamefont {Del~Pozzo},\ and\
  \citenamefont {Van Den~Broeck}}]{PhysRevD.98.104020}%
  \BibitemOpen
  \bibfield  {author} {\bibinfo {author} {\bibfnamefont {G.}~\bibnamefont
  {Carullo}}, \bibinfo {author} {\bibfnamefont {L.}~\bibnamefont {van~der
  Schaaf}}, \bibinfo {author} {\bibfnamefont {L.}~\bibnamefont {London}},
  \bibinfo {author} {\bibfnamefont {P.~T.~H.}\ \bibnamefont {Pang}}, \bibinfo
  {author} {\bibfnamefont {K.~W.}\ \bibnamefont {Tsang}}, \bibinfo {author}
  {\bibfnamefont {O.~A.}\ \bibnamefont {Hannuksela}}, \bibinfo {author}
  {\bibfnamefont {J.}~\bibnamefont {Meidam}}, \bibinfo {author} {\bibfnamefont
  {M.}~\bibnamefont {Agathos}}, \bibinfo {author} {\bibfnamefont
  {A.}~\bibnamefont {Samajdar}}, \bibinfo {author} {\bibfnamefont
  {A.}~\bibnamefont {Ghosh}}, \bibinfo {author} {\bibfnamefont {T.~G.~F.}\
  \bibnamefont {Li}}, \bibinfo {author} {\bibfnamefont {W.}~\bibnamefont
  {Del~Pozzo}}, \ and\ \bibinfo {author} {\bibfnamefont {C.}~\bibnamefont {Van
  Den~Broeck}},\ }\href {\doibase 10.1103/PhysRevD.98.104020} {\bibfield
  {journal} {\bibinfo  {journal} {Phys. Rev. D}\ }\textbf {\bibinfo {volume}
  {98}},\ \bibinfo {pages} {104020} (\bibinfo {year} {2018})}\BibitemShut
  {NoStop}%
\bibitem [{\citenamefont {Abbott}\ \emph {et~al.}(2016)\citenamefont {Abbott}
  \emph {et~al.}}]{TheLIGOScientific:2016src}%
  \BibitemOpen
  \bibfield  {author} {\bibinfo {author} {\bibfnamefont {B.~P.}\ \bibnamefont
  {Abbott}} \emph {et~al.} (\bibinfo {collaboration} {Virgo, LIGO
  Scientific}),\ }\href {\doibase 10.1103/PhysRevLett.116.221101,
  10.1103/PhysRevLett.121.129902} {\bibfield  {journal} {\bibinfo  {journal}
  {Phys. Rev. Lett.}\ }\textbf {\bibinfo {volume} {116}},\ \bibinfo {pages}
  {221101} (\bibinfo {year} {2016})},\ \bibinfo {note} {[Erratum: Phys. Rev.
  Lett.121,no.12,129902(2018)]},\ \Eprint {http://arxiv.org/abs/1602.03841}
  {arXiv:1602.03841 [gr-qc]} \BibitemShut {NoStop}%
\bibitem [{\citenamefont {Healy}\ \emph {et~al.}(2014)\citenamefont {Healy},
  \citenamefont {Lousto},\ and\ \citenamefont {Zlochower}}]{Healy:2014yta}%
  \BibitemOpen
  \bibfield  {author} {\bibinfo {author} {\bibfnamefont {J.}~\bibnamefont
  {Healy}}, \bibinfo {author} {\bibfnamefont {C.~O.}\ \bibnamefont {Lousto}}, \
  and\ \bibinfo {author} {\bibfnamefont {Y.}~\bibnamefont {Zlochower}},\ }\href
  {\doibase 10.1103/PhysRevD.90.104004} {\bibfield  {journal} {\bibinfo
  {journal} {Phys. Rev.}\ }\textbf {\bibinfo {volume} {D90}},\ \bibinfo {pages}
  {104004} (\bibinfo {year} {2014})},\ \Eprint {http://arxiv.org/abs/1406.7295}
  {arXiv:1406.7295 [gr-qc]} \BibitemShut {NoStop}%
\bibitem [{\citenamefont {Rezzolla}\ \emph {et~al.}(2008)\citenamefont
  {Rezzolla}, \citenamefont {Diener}, \citenamefont {Dorband}, \citenamefont
  {Pollney}, \citenamefont {Reisswig}, \citenamefont {Schnetter},\ and\
  \citenamefont {Seiler}}]{Rezzolla:2007rd}%
  \BibitemOpen
  \bibfield  {author} {\bibinfo {author} {\bibfnamefont {L.}~\bibnamefont
  {Rezzolla}}, \bibinfo {author} {\bibfnamefont {P.}~\bibnamefont {Diener}},
  \bibinfo {author} {\bibfnamefont {E.~N.}\ \bibnamefont {Dorband}}, \bibinfo
  {author} {\bibfnamefont {D.}~\bibnamefont {Pollney}}, \bibinfo {author}
  {\bibfnamefont {C.}~\bibnamefont {Reisswig}}, \bibinfo {author}
  {\bibfnamefont {E.}~\bibnamefont {Schnetter}}, \ and\ \bibinfo {author}
  {\bibfnamefont {J.}~\bibnamefont {Seiler}},\ }\href {\doibase 10.1086/528935}
  {\bibfield  {journal} {\bibinfo  {journal} {Astrophys. J.}\ }\textbf
  {\bibinfo {volume} {674}},\ \bibinfo {pages} {L29} (\bibinfo {year}
  {2008})},\ \Eprint {http://arxiv.org/abs/0710.3345} {arXiv:0710.3345 [gr-qc]}
  \BibitemShut {NoStop}%
\bibitem [{\citenamefont {{Healy}}\ and\ \citenamefont
  {{Lousto}}(2017)}]{2017PhRvD..95b4037H}%
  \BibitemOpen
  \bibfield  {author} {\bibinfo {author} {\bibfnamefont {J.}~\bibnamefont
  {{Healy}}}\ and\ \bibinfo {author} {\bibfnamefont {C.~O.}\ \bibnamefont
  {{Lousto}}},\ }\href {\doibase 10.1103/PhysRevD.95.024037} {\bibfield
  {journal} {\bibinfo  {journal} {PRD}\ }\textbf {\bibinfo {volume} {95}},\
  \bibinfo {eid} {024037} (\bibinfo {year} {2017})},\ \Eprint
  {http://arxiv.org/abs/1610.09713} {arXiv:1610.09713 [gr-qc]} \BibitemShut
  {NoStop}%
\bibitem [{\citenamefont {{Jim{\'e}nez-Forteza}}\ \emph
  {et~al.}(2017)\citenamefont {{Jim{\'e}nez-Forteza}}, \citenamefont
  {{Keitel}}, \citenamefont {{Husa}}, \citenamefont {{Hannam}}, \citenamefont
  {{Khan}},\ and\ \citenamefont {{P{\"u}rrer}}}]{2017PhRvD..95f4024J}%
  \BibitemOpen
  \bibfield  {author} {\bibinfo {author} {\bibfnamefont {X.}~\bibnamefont
  {{Jim{\'e}nez-Forteza}}}, \bibinfo {author} {\bibfnamefont {D.}~\bibnamefont
  {{Keitel}}}, \bibinfo {author} {\bibfnamefont {S.}~\bibnamefont {{Husa}}},
  \bibinfo {author} {\bibfnamefont {M.}~\bibnamefont {{Hannam}}}, \bibinfo
  {author} {\bibfnamefont {S.}~\bibnamefont {{Khan}}}, \ and\ \bibinfo {author}
  {\bibfnamefont {M.}~\bibnamefont {{P{\"u}rrer}}},\ }\href {\doibase
  10.1103/PhysRevD.95.064024} {\bibfield  {journal} {\bibinfo  {journal} {PRD}\
  }\textbf {\bibinfo {volume} {95}},\ \bibinfo {eid} {064024} (\bibinfo {year}
  {2017})},\ \Eprint {http://arxiv.org/abs/1611.00332} {arXiv:1611.00332
  [gr-qc]} \BibitemShut {NoStop}%
\bibitem [{\citenamefont {{Hofmann}}\ \emph {et~al.}(2016)\citenamefont
  {{Hofmann}}, \citenamefont {{Barausse}},\ and\ \citenamefont
  {{Rezzolla}}}]{2016ApJ...825L..19H}%
  \BibitemOpen
  \bibfield  {author} {\bibinfo {author} {\bibfnamefont {F.}~\bibnamefont
  {{Hofmann}}}, \bibinfo {author} {\bibfnamefont {E.}~\bibnamefont
  {{Barausse}}}, \ and\ \bibinfo {author} {\bibfnamefont {L.}~\bibnamefont
  {{Rezzolla}}},\ }\href {\doibase 10.3847/2041-8205/825/2/L19} {\bibfield
  {journal} {\bibinfo  {journal} {ApJL}\ }\textbf {\bibinfo {volume} {825}},\
  \bibinfo {eid} {L19} (\bibinfo {year} {2016})},\ \Eprint
  {http://arxiv.org/abs/1605.01938} {arXiv:1605.01938 [gr-qc]} \BibitemShut
  {NoStop}%
\bibitem [{\citenamefont {Abbott}\ \emph {et~al.}(2019)\citenamefont {Abbott}
  \emph {et~al.}}]{LIGOScientific:2019fpa}%
  \BibitemOpen
  \bibfield  {author} {\bibinfo {author} {\bibfnamefont {B.~P.}\ \bibnamefont
  {Abbott}} \emph {et~al.} (\bibinfo {collaboration} {LIGO Scientific,
  Virgo}),\ }\href@noop {} {\  (\bibinfo {year} {2019})},\ \Eprint
  {http://arxiv.org/abs/1903.04467} {arXiv:1903.04467 [gr-qc]} \BibitemShut
  {NoStop}%
\bibitem [{\citenamefont {Ghosh}\ \emph {et~al.}(2018)\citenamefont {Ghosh},
  \citenamefont {Johnson-Mcdaniel}, \citenamefont {Ghosh}, \citenamefont
  {Mishra}, \citenamefont {Ajith}, \citenamefont {Del~Pozzo}, \citenamefont
  {Berry}, \citenamefont {Nielsen},\ and\ \citenamefont
  {London}}]{Ghosh:2017gfp}%
  \BibitemOpen
  \bibfield  {author} {\bibinfo {author} {\bibfnamefont {A.}~\bibnamefont
  {Ghosh}}, \bibinfo {author} {\bibfnamefont {N.~K.}\ \bibnamefont
  {Johnson-Mcdaniel}}, \bibinfo {author} {\bibfnamefont {A.}~\bibnamefont
  {Ghosh}}, \bibinfo {author} {\bibfnamefont {C.~K.}\ \bibnamefont {Mishra}},
  \bibinfo {author} {\bibfnamefont {P.}~\bibnamefont {Ajith}}, \bibinfo
  {author} {\bibfnamefont {W.}~\bibnamefont {Del~Pozzo}}, \bibinfo {author}
  {\bibfnamefont {C.~P.~L.}\ \bibnamefont {Berry}}, \bibinfo {author}
  {\bibfnamefont {A.~B.}\ \bibnamefont {Nielsen}}, \ and\ \bibinfo {author}
  {\bibfnamefont {L.}~\bibnamefont {London}},\ }\href {\doibase
  10.1088/1361-6382/aa972e} {\bibfield  {journal} {\bibinfo  {journal} {Class.
  Quant. Grav.}\ }\textbf {\bibinfo {volume} {35}},\ \bibinfo {pages} {014002}
  (\bibinfo {year} {2018})},\ \Eprint {http://arxiv.org/abs/1704.06784}
  {arXiv:1704.06784 [gr-qc]} \BibitemShut {NoStop}%
\bibitem [{\citenamefont {Ghosh}\ \emph {et~al.}(2016)\citenamefont {Ghosh}
  \emph {et~al.}}]{Ghosh:2016qgn}%
  \BibitemOpen
  \bibfield  {author} {\bibinfo {author} {\bibfnamefont {A.}~\bibnamefont
  {Ghosh}} \emph {et~al.},\ }\href {\doibase 10.1103/PhysRevD.94.021101}
  {\bibfield  {journal} {\bibinfo  {journal} {Phys. Rev.}\ }\textbf {\bibinfo
  {volume} {D94}},\ \bibinfo {pages} {021101} (\bibinfo {year} {2016})},\
  \Eprint {http://arxiv.org/abs/1602.02453} {arXiv:1602.02453 [gr-qc]}
  \BibitemShut {NoStop}%
\bibitem [{\citenamefont {Carullo}\ \emph
  {et~al.}(2019{\natexlab{b}})\citenamefont {Carullo}, \citenamefont
  {Riemenschneider}, \citenamefont {Tsang}, \citenamefont {Nagar},\ and\
  \citenamefont {Del~Pozzo}}]{Carullo:2018gah}%
  \BibitemOpen
  \bibfield  {author} {\bibinfo {author} {\bibfnamefont {G.}~\bibnamefont
  {Carullo}}, \bibinfo {author} {\bibfnamefont {G.}~\bibnamefont
  {Riemenschneider}}, \bibinfo {author} {\bibfnamefont {K.~W.}\ \bibnamefont
  {Tsang}}, \bibinfo {author} {\bibfnamefont {A.}~\bibnamefont {Nagar}}, \ and\
  \bibinfo {author} {\bibfnamefont {W.}~\bibnamefont {Del~Pozzo}},\ }\href
  {\doibase 10.1088/1361-6382/ab185e} {\bibfield  {journal} {\bibinfo
  {journal} {Class. Quant. Grav.}\ }\textbf {\bibinfo {volume} {36}},\ \bibinfo
  {pages} {105009} (\bibinfo {year} {2019}{\natexlab{b}})},\ \Eprint
  {http://arxiv.org/abs/1811.08744} {arXiv:1811.08744 [gr-qc]} \BibitemShut
  {NoStop}%
\bibitem [{\citenamefont {{Healy}}\ \emph {et~al.}(2014)\citenamefont
  {{Healy}}, \citenamefont {{Laguna}},\ and\ \citenamefont
  {{Shoemaker}}}]{2014CQGra..31u2001H}%
  \BibitemOpen
  \bibfield  {author} {\bibinfo {author} {\bibfnamefont {J.}~\bibnamefont
  {{Healy}}}, \bibinfo {author} {\bibfnamefont {P.}~\bibnamefont {{Laguna}}}, \
  and\ \bibinfo {author} {\bibfnamefont {D.}~\bibnamefont {{Shoemaker}}},\
  }\href {\doibase 10.1088/0264-9381/31/21/212001} {\bibfield  {journal}
  {\bibinfo  {journal} {Classical and Quantum Gravity}\ }\textbf {\bibinfo
  {volume} {31}},\ \bibinfo {eid} {212001} (\bibinfo {year} {2014})},\ \Eprint
  {http://arxiv.org/abs/1407.5989} {arXiv:1407.5989 [gr-qc]} \BibitemShut
  {NoStop}%
\bibitem [{\citenamefont {Buonanno}\ \emph {et~al.}(2007)\citenamefont
  {Buonanno}, \citenamefont {Cook},\ and\ \citenamefont
  {Pretorius}}]{Buonanno:2006ui}%
  \BibitemOpen
  \bibfield  {author} {\bibinfo {author} {\bibfnamefont {A.}~\bibnamefont
  {Buonanno}}, \bibinfo {author} {\bibfnamefont {G.~B.}\ \bibnamefont {Cook}},
  \ and\ \bibinfo {author} {\bibfnamefont {F.}~\bibnamefont {Pretorius}},\
  }\href {\doibase 10.1103/PhysRevD.75.124018} {\bibfield  {journal} {\bibinfo
  {journal} {Phys. Rev.}\ }\textbf {\bibinfo {volume} {D75}},\ \bibinfo {pages}
  {124018} (\bibinfo {year} {2007})},\ \Eprint
  {http://arxiv.org/abs/gr-qc/0610122} {arXiv:gr-qc/0610122 [gr-qc]}
  \BibitemShut {NoStop}%
\bibitem [{\citenamefont {Kamaretsos}\ \emph {et~al.}(2012)\citenamefont
  {Kamaretsos}, \citenamefont {Hannam},\ and\ \citenamefont
  {Sathyaprakash}}]{PhysRevLett.109.141102}%
  \BibitemOpen
  \bibfield  {author} {\bibinfo {author} {\bibfnamefont {I.}~\bibnamefont
  {Kamaretsos}}, \bibinfo {author} {\bibfnamefont {M.}~\bibnamefont {Hannam}},
  \ and\ \bibinfo {author} {\bibfnamefont {B.~S.}\ \bibnamefont
  {Sathyaprakash}},\ }\href {\doibase 10.1103/PhysRevLett.109.141102}
  {\bibfield  {journal} {\bibinfo  {journal} {Phys. Rev. Lett.}\ }\textbf
  {\bibinfo {volume} {109}},\ \bibinfo {pages} {141102} (\bibinfo {year}
  {2012})}\BibitemShut {NoStop}%
\bibitem [{\citenamefont {Giesler}\ \emph {et~al.}(2019)\citenamefont
  {Giesler}, \citenamefont {Isi}, \citenamefont {Scheel},\ and\ \citenamefont
  {Teukolsky}}]{Giesler:2019uxc}%
  \BibitemOpen
  \bibfield  {author} {\bibinfo {author} {\bibfnamefont {M.}~\bibnamefont
  {Giesler}}, \bibinfo {author} {\bibfnamefont {M.}~\bibnamefont {Isi}},
  \bibinfo {author} {\bibfnamefont {M.}~\bibnamefont {Scheel}}, \ and\ \bibinfo
  {author} {\bibfnamefont {S.}~\bibnamefont {Teukolsky}},\ }\href@noop {} {\
  (\bibinfo {year} {2019})},\ \Eprint {http://arxiv.org/abs/1903.08284}
  {arXiv:1903.08284 [gr-qc]} \BibitemShut {NoStop}%
\bibitem [{\citenamefont {Tiwari}\ \emph {et~al.}(2016)\citenamefont {Tiwari},
  \citenamefont {Klimenko}, \citenamefont {Necula},\ and\ \citenamefont
  {Mitselmakher}}]{Tiwari:2015bda}%
  \BibitemOpen
  \bibfield  {author} {\bibinfo {author} {\bibfnamefont {V.}~\bibnamefont
  {Tiwari}}, \bibinfo {author} {\bibfnamefont {S.}~\bibnamefont {Klimenko}},
  \bibinfo {author} {\bibfnamefont {V.}~\bibnamefont {Necula}}, \ and\ \bibinfo
  {author} {\bibfnamefont {G.}~\bibnamefont {Mitselmakher}},\ }\href {\doibase
  10.1088/0264-9381/33/1/01LT01} {\bibfield  {journal} {\bibinfo  {journal}
  {Class. Quant. Grav.}\ }\textbf {\bibinfo {volume} {33}},\ \bibinfo {pages}
  {01LT01} (\bibinfo {year} {2016})},\ \Eprint
  {http://arxiv.org/abs/1510.02426} {arXiv:1510.02426 [astro-ph.IM]}
  \BibitemShut {NoStop}%
\bibitem [{\citenamefont {{Jani}}\ \emph {et~al.}(2016)\citenamefont {{Jani}},
  \citenamefont {{Healy}}, \citenamefont {{Clark}}, \citenamefont {{London}},
  \citenamefont {{Laguna}},\ and\ \citenamefont
  {{Shoemaker}}}]{2016CQGra..33t4001J}%
  \BibitemOpen
  \bibfield  {author} {\bibinfo {author} {\bibfnamefont {K.}~\bibnamefont
  {{Jani}}}, \bibinfo {author} {\bibfnamefont {J.}~\bibnamefont {{Healy}}},
  \bibinfo {author} {\bibfnamefont {J.~A.}\ \bibnamefont {{Clark}}}, \bibinfo
  {author} {\bibfnamefont {L.}~\bibnamefont {{London}}}, \bibinfo {author}
  {\bibfnamefont {P.}~\bibnamefont {{Laguna}}}, \ and\ \bibinfo {author}
  {\bibfnamefont {D.}~\bibnamefont {{Shoemaker}}},\ }\href {\doibase
  10.1088/0264-9381/33/20/204001} {\bibfield  {journal} {\bibinfo  {journal}
  {Classical and Quantum Gravity}\ }\textbf {\bibinfo {volume} {33}},\ \bibinfo
  {eid} {204001} (\bibinfo {year} {2016})},\ \Eprint
  {http://arxiv.org/abs/1605.03204} {arXiv:1605.03204 [gr-qc]} \BibitemShut
  {NoStop}%
\bibitem [{\citenamefont {{Herrmann}}\ \emph {et~al.}(2007)\citenamefont
  {{Herrmann}}, \citenamefont {{Hinder}}, \citenamefont {{Shoemaker}},\ and\
  \citenamefont {{Laguna}}}]{2007CQGra..24S..33H}%
  \BibitemOpen
  \bibfield  {author} {\bibinfo {author} {\bibfnamefont {F.}~\bibnamefont
  {{Herrmann}}}, \bibinfo {author} {\bibfnamefont {I.}~\bibnamefont
  {{Hinder}}}, \bibinfo {author} {\bibfnamefont {D.}~\bibnamefont
  {{Shoemaker}}}, \ and\ \bibinfo {author} {\bibfnamefont {P.}~\bibnamefont
  {{Laguna}}},\ }\href {\doibase 10.1088/0264-9381/24/12/S04} {\bibfield
  {journal} {\bibinfo  {journal} {Classical and Quantum Gravity}\ }\textbf
  {\bibinfo {volume} {24}},\ \bibinfo {pages} {S33} (\bibinfo {year} {2007})},\
  \Eprint {http://arxiv.org/abs/gr-qc/0601026} {arXiv:gr-qc/0601026 [gr-qc]}
  \BibitemShut {NoStop}%
\bibitem [{\citenamefont {{Vaishnav}}\ \emph {et~al.}(2007)\citenamefont
  {{Vaishnav}}, \citenamefont {{Hinder}}, \citenamefont {{Herrmann}},\ and\
  \citenamefont {{Shoemaker}}}]{2007PhRvD..76h4020V}%
  \BibitemOpen
  \bibfield  {author} {\bibinfo {author} {\bibfnamefont {B.}~\bibnamefont
  {{Vaishnav}}}, \bibinfo {author} {\bibfnamefont {I.}~\bibnamefont
  {{Hinder}}}, \bibinfo {author} {\bibfnamefont {F.}~\bibnamefont
  {{Herrmann}}}, \ and\ \bibinfo {author} {\bibfnamefont {D.}~\bibnamefont
  {{Shoemaker}}},\ }\href {\doibase 10.1103/PhysRevD.76.084020} {\bibfield
  {journal} {\bibinfo  {journal} {\prd}\ }\textbf {\bibinfo {volume} {76}},\
  \bibinfo {eid} {084020} (\bibinfo {year} {2007})},\ \Eprint
  {http://arxiv.org/abs/0705.3829} {arXiv:0705.3829 [gr-qc]} \BibitemShut
  {NoStop}%
\bibitem [{\citenamefont {Healy}\ \emph {et~al.}(2009)\citenamefont {Healy},
  \citenamefont {Levin},\ and\ \citenamefont
  {Shoemaker}}]{PhysRevLett.103.131101}%
  \BibitemOpen
  \bibfield  {author} {\bibinfo {author} {\bibfnamefont {J.}~\bibnamefont
  {Healy}}, \bibinfo {author} {\bibfnamefont {J.}~\bibnamefont {Levin}}, \ and\
  \bibinfo {author} {\bibfnamefont {D.}~\bibnamefont {Shoemaker}},\ }\href
  {\doibase 10.1103/PhysRevLett.103.131101} {\bibfield  {journal} {\bibinfo
  {journal} {Phys. Rev. Lett.}\ }\textbf {\bibinfo {volume} {103}},\ \bibinfo
  {pages} {131101} (\bibinfo {year} {2009})}\BibitemShut {NoStop}%
\bibitem [{\citenamefont {Pekowsky}\ \emph
  {et~al.}(2013{\natexlab{a}})\citenamefont {Pekowsky}, \citenamefont
  {O'Shaughnessy}, \citenamefont {Healy},\ and\ \citenamefont
  {Shoemaker}}]{PhysRevD.88.024040}%
  \BibitemOpen
  \bibfield  {author} {\bibinfo {author} {\bibfnamefont {L.}~\bibnamefont
  {Pekowsky}}, \bibinfo {author} {\bibfnamefont {R.}~\bibnamefont
  {O'Shaughnessy}}, \bibinfo {author} {\bibfnamefont {J.}~\bibnamefont
  {Healy}}, \ and\ \bibinfo {author} {\bibfnamefont {D.}~\bibnamefont
  {Shoemaker}},\ }\href {\doibase 10.1103/PhysRevD.88.024040} {\bibfield
  {journal} {\bibinfo  {journal} {Phys. Rev. D}\ }\textbf {\bibinfo {volume}
  {88}},\ \bibinfo {pages} {024040} (\bibinfo {year}
  {2013}{\natexlab{a}})}\BibitemShut {NoStop}%
\bibitem [{\citenamefont {Loffler}\ \emph {et~al.}(2012)\citenamefont {Loffler}
  \emph {et~al.}}]{Loffler:2011ay}%
  \BibitemOpen
  \bibfield  {author} {\bibinfo {author} {\bibfnamefont {F.}~\bibnamefont
  {Loffler}} \emph {et~al.},\ }\href {\doibase 10.1088/0264-9381/29/11/115001}
  {\bibfield  {journal} {\bibinfo  {journal} {Class. Quant. Grav.}\ }\textbf
  {\bibinfo {volume} {29}},\ \bibinfo {pages} {115001} (\bibinfo {year}
  {2012})},\ \Eprint {http://arxiv.org/abs/1111.3344} {arXiv:1111.3344 [gr-qc]}
  \BibitemShut {NoStop}%
\bibitem [{\citenamefont {Schnetter}\ \emph {et~al.}(2004)\citenamefont
  {Schnetter}, \citenamefont {Hawley},\ and\ \citenamefont
  {Hawke}}]{Schnetter:2003rb}%
  \BibitemOpen
  \bibfield  {author} {\bibinfo {author} {\bibfnamefont {E.}~\bibnamefont
  {Schnetter}}, \bibinfo {author} {\bibfnamefont {S.~H.}\ \bibnamefont
  {Hawley}}, \ and\ \bibinfo {author} {\bibfnamefont {I.}~\bibnamefont
  {Hawke}},\ }\href {\doibase 10.1088/0264-9381/21/6/014} {\bibfield  {journal}
  {\bibinfo  {journal} {Class. Quant. Grav.}\ }\textbf {\bibinfo {volume}
  {21}},\ \bibinfo {pages} {1465} (\bibinfo {year} {2004})},\ \Eprint
  {http://arxiv.org/abs/gr-qc/0310042} {arXiv:gr-qc/0310042 [gr-qc]}
  \BibitemShut {NoStop}%
\bibitem [{\citenamefont {Thornburg}(2004)}]{Thornburg:2003sf}%
  \BibitemOpen
  \bibfield  {author} {\bibinfo {author} {\bibfnamefont {J.}~\bibnamefont
  {Thornburg}},\ }\href {\doibase 10.1088/0264-9381/21/2/026} {\bibfield
  {journal} {\bibinfo  {journal} {Class. Quant. Grav.}\ }\textbf {\bibinfo
  {volume} {21}},\ \bibinfo {pages} {743} (\bibinfo {year} {2004})},\ \Eprint
  {http://arxiv.org/abs/gr-qc/0306056} {arXiv:gr-qc/0306056 [gr-qc]}
  \BibitemShut {NoStop}%
\bibitem [{\citenamefont {Reisswig}\ and\ \citenamefont
  {Pollney}(2011)}]{Reisswig:2010di}%
  \BibitemOpen
  \bibfield  {author} {\bibinfo {author} {\bibfnamefont {C.}~\bibnamefont
  {Reisswig}}\ and\ \bibinfo {author} {\bibfnamefont {D.}~\bibnamefont
  {Pollney}},\ }\href {\doibase 10.1088/0264-9381/28/19/195015} {\bibfield
  {journal} {\bibinfo  {journal} {Class. Quant. Grav.}\ }\textbf {\bibinfo
  {volume} {28}},\ \bibinfo {pages} {195015} (\bibinfo {year} {2011})},\
  \Eprint {http://arxiv.org/abs/1006.1632} {arXiv:1006.1632 [gr-qc]}
  \BibitemShut {NoStop}%
\bibitem [{\citenamefont {Calderón~Bustillo}\ \emph
  {et~al.}(2017)\citenamefont {Calderón~Bustillo}, \citenamefont {Laguna},\
  and\ \citenamefont {Shoemaker}}]{Bustillo:2016gid}%
  \BibitemOpen
  \bibfield  {author} {\bibinfo {author} {\bibfnamefont {J.}~\bibnamefont
  {Calderón~Bustillo}}, \bibinfo {author} {\bibfnamefont {P.}~\bibnamefont
  {Laguna}}, \ and\ \bibinfo {author} {\bibfnamefont {D.}~\bibnamefont
  {Shoemaker}},\ }\href {\doibase 10.1103/PhysRevD.95.104038} {\bibfield
  {journal} {\bibinfo  {journal} {Phys. Rev.}\ }\textbf {\bibinfo {volume}
  {D95}},\ \bibinfo {pages} {104038} (\bibinfo {year} {2017})},\ \Eprint
  {http://arxiv.org/abs/1612.02340} {arXiv:1612.02340 [gr-qc]} \BibitemShut
  {NoStop}%
\bibitem [{\citenamefont {Calderón~Bustillo}\ \emph
  {et~al.}(2016)\citenamefont {Calderón~Bustillo}, \citenamefont {Husa},
  \citenamefont {Sintes},\ and\ \citenamefont {Pürrer}}]{Bustillo:2015qty}%
  \BibitemOpen
  \bibfield  {author} {\bibinfo {author} {\bibfnamefont {J.}~\bibnamefont
  {Calderón~Bustillo}}, \bibinfo {author} {\bibfnamefont {S.}~\bibnamefont
  {Husa}}, \bibinfo {author} {\bibfnamefont {A.~M.}\ \bibnamefont {Sintes}}, \
  and\ \bibinfo {author} {\bibfnamefont {M.}~\bibnamefont {Pürrer}},\ }\href
  {\doibase 10.1103/PhysRevD.93.084019} {\bibfield  {journal} {\bibinfo
  {journal} {Phys. Rev.}\ }\textbf {\bibinfo {volume} {D93}},\ \bibinfo {pages}
  {084019} (\bibinfo {year} {2016})},\ \Eprint
  {http://arxiv.org/abs/1511.02060} {arXiv:1511.02060 [gr-qc]} \BibitemShut
  {NoStop}%
\bibitem [{\citenamefont {Pekowsky}\ \emph
  {et~al.}(2013{\natexlab{b}})\citenamefont {Pekowsky}, \citenamefont {Healy},
  \citenamefont {Shoemaker},\ and\ \citenamefont {Laguna}}]{Pekowsky:2012sr}%
  \BibitemOpen
  \bibfield  {author} {\bibinfo {author} {\bibfnamefont {L.}~\bibnamefont
  {Pekowsky}}, \bibinfo {author} {\bibfnamefont {J.}~\bibnamefont {Healy}},
  \bibinfo {author} {\bibfnamefont {D.}~\bibnamefont {Shoemaker}}, \ and\
  \bibinfo {author} {\bibfnamefont {P.}~\bibnamefont {Laguna}},\ }\href
  {\doibase 10.1103/PhysRevD.87.084008} {\bibfield  {journal} {\bibinfo
  {journal} {Phys. Rev.}\ }\textbf {\bibinfo {volume} {D87}},\ \bibinfo {pages}
  {084008} (\bibinfo {year} {2013}{\natexlab{b}})},\ \Eprint
  {http://arxiv.org/abs/1210.1891} {arXiv:1210.1891 [gr-qc]} \BibitemShut
  {NoStop}%
\bibitem [{\citenamefont {Varma}\ and\ \citenamefont
  {Ajith}(2017)}]{Varma:2016dnf}%
  \BibitemOpen
  \bibfield  {author} {\bibinfo {author} {\bibfnamefont {V.}~\bibnamefont
  {Varma}}\ and\ \bibinfo {author} {\bibfnamefont {P.}~\bibnamefont {Ajith}},\
  }\href {\doibase 10.1103/PhysRevD.96.124024} {\bibfield  {journal} {\bibinfo
  {journal} {Phys. Rev.}\ }\textbf {\bibinfo {volume} {D96}},\ \bibinfo {pages}
  {124024} (\bibinfo {year} {2017})},\ \Eprint
  {http://arxiv.org/abs/1612.05608} {arXiv:1612.05608 [gr-qc]} \BibitemShut
  {NoStop}%
\bibitem [{\citenamefont {Vallisneri}(2008)}]{Vallisneri:2007ev}%
  \BibitemOpen
  \bibfield  {author} {\bibinfo {author} {\bibfnamefont {M.}~\bibnamefont
  {Vallisneri}},\ }\href {\doibase 10.1103/PhysRevD.77.042001} {\bibfield
  {journal} {\bibinfo  {journal} {Phys. Rev.}\ }\textbf {\bibinfo {volume}
  {D77}},\ \bibinfo {pages} {042001} (\bibinfo {year} {2008})},\ \Eprint
  {http://arxiv.org/abs/gr-qc/0703086} {arXiv:gr-qc/0703086 [GR-QC]}
  \BibitemShut {NoStop}%
\bibitem [{\citenamefont {Cornish}\ and\ \citenamefont
  {Littenberg}(2015)}]{Cornish:2014kda}%
  \BibitemOpen
  \bibfield  {author} {\bibinfo {author} {\bibfnamefont {N.~J.}\ \bibnamefont
  {Cornish}}\ and\ \bibinfo {author} {\bibfnamefont {T.~B.}\ \bibnamefont
  {Littenberg}},\ }\href {\doibase 10.1088/0264-9381/32/13/135012} {\bibfield
  {journal} {\bibinfo  {journal} {Class. Quant. Grav.}\ }\textbf {\bibinfo
  {volume} {32}},\ \bibinfo {pages} {135012} (\bibinfo {year} {2015})},\
  \Eprint {http://arxiv.org/abs/1410.3835} {arXiv:1410.3835 [gr-qc]}
  \BibitemShut {NoStop}%
\bibitem [{\citenamefont {Fairhurst}\ and\ \citenamefont
  {Brady}(2008)}]{Fairhurst:2007qj}%
  \BibitemOpen
  \bibfield  {author} {\bibinfo {author} {\bibfnamefont {S.}~\bibnamefont
  {Fairhurst}}\ and\ \bibinfo {author} {\bibfnamefont {P.}~\bibnamefont
  {Brady}},\ }\href {\doibase 10.1088/0264-9381/25/10/105002} {\bibfield
  {journal} {\bibinfo  {journal} {Class. Quant. Grav.}\ }\textbf {\bibinfo
  {volume} {25}},\ \bibinfo {pages} {105002} (\bibinfo {year}
  {2008})}\BibitemShut {NoStop}%
\bibitem [{\citenamefont {Veitch}\ \emph {et~al.}(2015)\citenamefont {Veitch},
  \citenamefont {Raymond}, \citenamefont {Farr}, \citenamefont {Farr},
  \citenamefont {Graff}, \citenamefont {Vitale}, \citenamefont {Aylott},
  \citenamefont {Blackburn}, \citenamefont {Christensen}, \citenamefont
  {Coughlin} \emph {et~al.}}]{veitch2015parameter}%
  \BibitemOpen
  \bibfield  {author} {\bibinfo {author} {\bibfnamefont {J.}~\bibnamefont
  {Veitch}}, \bibinfo {author} {\bibfnamefont {V.}~\bibnamefont {Raymond}},
  \bibinfo {author} {\bibfnamefont {B.}~\bibnamefont {Farr}}, \bibinfo {author}
  {\bibfnamefont {W.}~\bibnamefont {Farr}}, \bibinfo {author} {\bibfnamefont
  {P.}~\bibnamefont {Graff}}, \bibinfo {author} {\bibfnamefont
  {S.}~\bibnamefont {Vitale}}, \bibinfo {author} {\bibfnamefont
  {B.}~\bibnamefont {Aylott}}, \bibinfo {author} {\bibfnamefont
  {K.}~\bibnamefont {Blackburn}}, \bibinfo {author} {\bibfnamefont
  {N.}~\bibnamefont {Christensen}}, \bibinfo {author} {\bibfnamefont
  {M.}~\bibnamefont {Coughlin}},  \emph {et~al.},\ }\href@noop {} {\bibfield
  {journal} {\bibinfo  {journal} {Physical Review D}\ }\textbf {\bibinfo
  {volume} {91}},\ \bibinfo {pages} {042003} (\bibinfo {year}
  {2015})}\BibitemShut {NoStop}%
\bibitem [{\citenamefont {{Messick}}\ \emph {et~al.}(2017)\citenamefont
  {{Messick}}, \citenamefont {{Blackburn}}, \citenamefont {{Brady}},
  \citenamefont {{Brockill}}, \citenamefont {{Cannon}}, \citenamefont
  {{Cariou}}, \citenamefont {{Caudill}}, \citenamefont {{Chamberlin}},
  \citenamefont {{Creighton}}, \citenamefont {{Everett}}, \citenamefont
  {{Hanna}}, \citenamefont {{Keppel}}, \citenamefont {{Lang}}, \citenamefont
  {{Li}}, \citenamefont {{Meacher}}, \citenamefont {{Nielsen}}, \citenamefont
  {{Pankow}}, \citenamefont {{Privitera}}, \citenamefont {{Qi}}, \citenamefont
  {{Sachdev}}, \citenamefont {{Sadeghian}}, \citenamefont {{Singer}},
  \citenamefont {{Thomas}}, \citenamefont {{Wade}}, \citenamefont {{Wade}},
  \citenamefont {{Weinstein}},\ and\ \citenamefont
  {{Wiesner}}}]{2017PhRvD..95d2001M}%
  \BibitemOpen
  \bibfield  {author} {\bibinfo {author} {\bibfnamefont {C.}~\bibnamefont
  {{Messick}}}, \bibinfo {author} {\bibfnamefont {K.}~\bibnamefont
  {{Blackburn}}}, \bibinfo {author} {\bibfnamefont {P.}~\bibnamefont
  {{Brady}}}, \bibinfo {author} {\bibfnamefont {P.}~\bibnamefont {{Brockill}}},
  \bibinfo {author} {\bibfnamefont {K.}~\bibnamefont {{Cannon}}}, \bibinfo
  {author} {\bibfnamefont {R.}~\bibnamefont {{Cariou}}}, \bibinfo {author}
  {\bibfnamefont {S.}~\bibnamefont {{Caudill}}}, \bibinfo {author}
  {\bibfnamefont {S.~J.}\ \bibnamefont {{Chamberlin}}}, \bibinfo {author}
  {\bibfnamefont {J.~D.~E.}\ \bibnamefont {{Creighton}}}, \bibinfo {author}
  {\bibfnamefont {R.}~\bibnamefont {{Everett}}}, \bibinfo {author}
  {\bibfnamefont {C.}~\bibnamefont {{Hanna}}}, \bibinfo {author} {\bibfnamefont
  {D.}~\bibnamefont {{Keppel}}}, \bibinfo {author} {\bibfnamefont {R.~N.}\
  \bibnamefont {{Lang}}}, \bibinfo {author} {\bibfnamefont {T.~G.~F.}\
  \bibnamefont {{Li}}}, \bibinfo {author} {\bibfnamefont {D.}~\bibnamefont
  {{Meacher}}}, \bibinfo {author} {\bibfnamefont {A.}~\bibnamefont
  {{Nielsen}}}, \bibinfo {author} {\bibfnamefont {C.}~\bibnamefont {{Pankow}}},
  \bibinfo {author} {\bibfnamefont {S.}~\bibnamefont {{Privitera}}}, \bibinfo
  {author} {\bibfnamefont {H.}~\bibnamefont {{Qi}}}, \bibinfo {author}
  {\bibfnamefont {S.}~\bibnamefont {{Sachdev}}}, \bibinfo {author}
  {\bibfnamefont {L.}~\bibnamefont {{Sadeghian}}}, \bibinfo {author}
  {\bibfnamefont {L.}~\bibnamefont {{Singer}}}, \bibinfo {author}
  {\bibfnamefont {E.~G.}\ \bibnamefont {{Thomas}}}, \bibinfo {author}
  {\bibfnamefont {L.}~\bibnamefont {{Wade}}}, \bibinfo {author} {\bibfnamefont
  {M.}~\bibnamefont {{Wade}}}, \bibinfo {author} {\bibfnamefont
  {A.}~\bibnamefont {{Weinstein}}}, \ and\ \bibinfo {author} {\bibfnamefont
  {K.}~\bibnamefont {{Wiesner}}},\ }\href {\doibase 10.1103/PhysRevD.95.042001}
  {\bibfield  {journal} {\bibinfo  {journal} {\prd}\ }\textbf {\bibinfo
  {volume} {95}},\ \bibinfo {eid} {042001} (\bibinfo {year} {2017})},\ \Eprint
  {http://arxiv.org/abs/1604.04324} {arXiv:1604.04324 [astro-ph.IM]}
  \BibitemShut {NoStop}%
\bibitem [{\citenamefont {Parzen}(2006)}]{Parzen:2006:RES:2263383.2270618}%
  \BibitemOpen
  \bibfield  {author} {\bibinfo {author} {\bibfnamefont {E.}~\bibnamefont
  {Parzen}},\ }\href {\doibase 10.1109/TIT.1963.1057829} {\bibfield  {journal}
  {\bibinfo  {journal} {IEEE Trans. Inf. Theor.}\ }\textbf {\bibinfo {volume}
  {9}},\ \bibinfo {pages} {127} (\bibinfo {year} {2006})}\BibitemShut {NoStop}%
\bibitem [{\citenamefont {Huang}\ \emph {et~al.}(1998)\citenamefont {Huang},
  \citenamefont {Shen}, \citenamefont {Long}, \citenamefont {Wu}, \citenamefont
  {Shih}, \citenamefont {Zheng}, \citenamefont {Yen}, \citenamefont {Tung},\
  and\ \citenamefont {Liu}}]{doi:10.1098/rspa.1998.0193}%
  \BibitemOpen
  \bibfield  {author} {\bibinfo {author} {\bibfnamefont {N.~E.}\ \bibnamefont
  {Huang}}, \bibinfo {author} {\bibfnamefont {Z.}~\bibnamefont {Shen}},
  \bibinfo {author} {\bibfnamefont {S.~R.}\ \bibnamefont {Long}}, \bibinfo
  {author} {\bibfnamefont {M.~C.}\ \bibnamefont {Wu}}, \bibinfo {author}
  {\bibfnamefont {H.~H.}\ \bibnamefont {Shih}}, \bibinfo {author}
  {\bibfnamefont {Q.}~\bibnamefont {Zheng}}, \bibinfo {author} {\bibfnamefont
  {N.-C.}\ \bibnamefont {Yen}}, \bibinfo {author} {\bibfnamefont {C.~C.}\
  \bibnamefont {Tung}}, \ and\ \bibinfo {author} {\bibfnamefont {H.~H.}\
  \bibnamefont {Liu}},\ }\href {\doibase 10.1098/rspa.1998.0193} {\bibfield
  {journal} {\bibinfo  {journal} {Proceedings of the Royal Society of London.
  Series A: Mathematical, Physical and Engineering Sciences}\ }\textbf
  {\bibinfo {volume} {454}},\ \bibinfo {pages} {903} (\bibinfo {year}
  {1998})},\ \Eprint
  {http://arxiv.org/abs/https://royalsocietypublishing.org/doi/pdf/10.1098/rspa.1998.0193}
  {https://royalsocietypublishing.org/doi/pdf/10.1098/rspa.1998.0193}
  \BibitemShut {NoStop}%
\end{thebibliography}%
\end{document}